\documentclass[a4paper,11pt]{article}
\pdfoutput=1 

\usepackage{amsmath,amsfonts,amssymb,graphicx,color}
\usepackage{longtable,booktabs}
\usepackage{placeins}

\AtBeginDocument{
\heavyrulewidth=.08em
\lightrulewidth=.05em
\cmidrulewidth=.03em
\belowrulesep=.65ex
\belowbottomsep=0pt
\aboverulesep=.4ex
\abovetopsep=0pt
\cmidrulesep=\doublerulesep
\cmidrulekern=.5em
\defaultaddspace=.5em
}

\usepackage{geometry}
\usepackage{setspace}
\usepackage{changepage}

\usepackage{jcappub} 

\usepackage{orcidlink}
\usepackage[T1]{fontenc} 
                     
\usepackage{multirow}      
\usepackage[bf]{caption}

\usepackage{subcaption}

\usepackage{blkarray, bigstrut}
\usepackage{xparse}

\usepackage{lipsum}  
\usepackage{graphbox}   
\usepackage{stackengine}

\usepackage{graphicx}
\usepackage{dcolumn}
\usepackage{hyperref}
\hypersetup{colorlinks}
\usepackage[T1]{fontenc} 
\usepackage[normalem]{ulem}
\usepackage[utf8]{inputenc}

\usepackage[nameinlink,noabbrev]{cleveref}
\crefname{section}{Section}{Sections}
\crefname{table}{Table}{Tables}
\crefname{appendix}{Appendix}{Appendices}
\Crefname{figure}{Figure}{Figures}
\Crefname{equation}{Equation}{Equations}
\Crefname{section}{Section}{Sections}
\Crefname{table}{Table}{Tables}

\usepackage{booktabs}
\usepackage{float} 

\usepackage{soul}

\usepackage{bm}
\let\vec\bm

\newcommand{\hmpc}{\,h^{-1}\text{Mpc}}
\newcommand{\hmpci}{\,h \, \text{Mpc}^{-1}}

\newcommand{\mI}{\mathcal{I}}
\newcommand{\mJ}{\mathcal{J}}

\newcommand{\vk}{\vec k}

\newcommand{\vx}{\vec x}

\newcommand{\vhn}{\hat{\vec n}}

\newcommand{\dD}{\delta_\text{D}}

\newcommand{\vt}{\boldsymbol\theta}
\newcommand{\vell}{\boldsymbol\ell}

\newcommand{\vnu}{\boldsymbol\nu}

\usepackage{float} 

\usepackage{booktabs}

\newcommand{\revised}[1]{#1}

\title{Modeling the 3-point correlation function of projected scalar fields on the sphere}

\author[a,b]{{Abraham Arvizu}\orcidlink{0009-0006-6538-2312},}

\author[b]{{Alejandro Aviles}\orcidlink{0000-0001-5998-3986},}
\emailAdd{aviles@icf.unam.mx}

\author[b]{{Juan Carlos Hidalgo}\orcidlink{0000-0001-9715-1232},}

\author[a]{{Eladio Moreno}\orcidlink{0000-0002-5400-2584},}

\author[a]{{Gustavo Niz}\orcidlink{0000-0002-1544-8946},}

\author[c]{{Mario A. Rodriguez-Meza}\orcidlink{0000-0003-1160-1488},}

\author[b]{{Sofía Samario}\orcidlink{0009-0001-5224-9153},}

\author{The LSST Dark Energy Science Collaboration}

\affiliation[a]{Departamento de Ciencias e Ingenierias, Universidad de Guanajuato, 37150, Le\'on, Guanajuato}

\affiliation[b]{Instituto de Ciencias F\'isicas, Universidad Nacional
Autónoma de México,  62210, Cuernavaca, Morelos.}

\affiliation[c]{Departamento de F\'isica, Instituto Nacional de Investigaciones Nucleares,
Apartado Postal 18-1027, Col. Escand\'on, Ciudad de M\'exico,11801, M\'exico.}

\keywords{Large-Scale Structure}

\abstract{
One of the main obstacles for the signal extraction of the three point correlation function using photometric surveys, such as the Rubin Observatory Legacy Survey of Space and Time (LSST), will be the prohibitive computation time required for dealing with a vast quantity of sources. Brute force algorithms, which naively scales as $\mathcal{O}(N^3)$ with the number of objects, can be further improved with tree methods but not enough to deal with large scale correlations of Rubin’s data. However, a harmonic basis decomposition of these higher order statistics reduces the time dramatically, to scale as a two-point correlation function with the number of objects, so that the signal can be extracted in a reasonable amount of time. In this work, we aim to develop the framework to use these expansions within the Limber approximation for scalar (or spin-0) fields, such as galaxy counts, weak lensing convergence or aperture masses. We develop an estimator to extract the signal from catalogs and different phenomenological and theoretical models for its description. The latter includes halo model and standard perturbation theory, to which we add a simple effective field theory prescription based on the short range of non-locality of cosmic fields, significantly improving the agreement with simulated data. In parallel to the modeling of the signal, we develop a code that can efficiently calculate three points correlations of more than 200 million data points (a full sky simulation with Nside=4096) in $\sim$40 minutes, or even less than 10 minutes using an approximation in the searching algorithm, on a single high-performance computing node, enabling a feasible analysis for the upcoming LSST data.
}

\begin{document} 
\maketitle
\flushbottom

\begin{section}{Introduction}




According to the inflationary paradigm, primordial fluctuations of the cosmic fields are expected to be drawn from \revised{Gaussian random field}, with non-Gaussian corrections well suppressed at cosmological scales. Such primordial fluctuations seed the matter clustering, which is well understood at early times when overdensities are small and the linearized gravity equations are reliable.
These two physical ingredients together imply that all the statistical information of the early Universe is mostly captured by the 2-point correlation function (2PCF).  
This early Gaussian picture is supported by observations of the primary anisotropies of the cosmic microwave background radiation (CMB) \cite{Planck:2019kim} as well as other Large Scale Structure (LSS) results, e.g.\cite{Slosar:2008hx,eBOSS:2019sma,eBOSS:2021jbt,DAmico:2022gki}. As the Universe evolves, however, non-linearities grow due to the gravitational collapse of cosmic structures. When this happens, a non-trivial signal appears on higher (than two) {\it N}-point correlation functions, which may get further enhanced by the growth of primordial non-Gaussianities, or
the presence of new physics in the gravitational sector \cite{Gil-Marin:2011iuh,Yamauchi:2017ibz,Aviles:2023fqx,Bose:2018zpk} or the matter sector \cite{Takada:2003ef,Bose:2018zpk}. 
Conversely, understanding the structure of higher order point correlation functions in the late time Universe may help disentangling any of the previous possibilities, while also helping to break degeneracies with systematics associated with observations and their analysis. 
Furthermore, higher order correlations may also shed light on properties of the matter distribution that cannot be inferred with the 2PCF. An example of this is the recent work that hints to a potential violation of parity in the three-dimensional distribution of galaxies using the four point correlation function \cite{Cahn:2021ltp,Philcox:2022hkh,Hou:2022wfj}. Although this parity violation may not be of primordial origin, since it is not present in the CMB temperature map \cite{Philcox:2023ffy}, the Lyman alpha forest \cite{Adari:2024vkf} or with a better estimation of the convariances \cite{Philcox:2024mmz, Krolewski:2024paz}, further studies are needed to fully characterise the signal. Similar ideas have been put forward for the bispectrum and other statistics beyond the 2PCF (see for example \cite{Sugiyama:2018yzo, Sugiyama:2017ggb, Wang:2023zkv}).

With the advent of a large wealth of data coming from \revised{Stage III and Stage IV surveys} and CMB probes such as the Hyper-Suprime Cam Survey\footnote{\url{https://hsc.mtk.nao.ac.jp/}} (HSC), the Dark Energy Survey\footnote{\url{https://www.darkenergysurvey.org/}} (DES), the Rubin Observatory Legacy Survey of Space and Time\footnote{\url{https://www.lsst.org/}} (LSST), Kilo Degree Survey\footnote{\url{https://kids.strw.leidenuniv.nl}} (KiDS), the Dark Energy Spectroscopic Instrument\footnote{\url{https://www.desi.lbl.gov/}} (DESI), Euclid\footnote{\url{https://www.euclid-ec.org/}}, the Keck/BICEP array\footnote{\url{array http://bicepkeck.org/}}, Adv-ACT\footnote{\url{https://act.princeton.edu/}}, the Roman Space Telescope \footnote{\url{https://roman.gsfc.nasa.gov/}}, the Simons Observatory\footnote{\url{https://simonsobservatory.org/}}, and CMB-S4\footnote{\url{https://cmb-s4.org/}}, the 3-point correlation function (3PCF), and potentially higher order correlations (HOC), will take more relevance as complementary statistical tools to the standard analyses based on two-point statistics \cite{Gualdi:2021yvq}. For this reason, novel and accurate descriptions of such higher order summary statistics can be proved useful to analyze current and upcoming data. The studies in \cite{Gil-Marin:2014sta,Hahn:2019zob,Brown:2024dmv} are a few examples of promising HOC research in galaxy surveys. However, a robust HOC pipeline would not only need the analytical or emphirical model of the signal for particular estimators (e.g. \cite{Takahashi:2019hth, Philcox:2022frc}), but also an estimation of the covariance matrices and understanding the effect of systematics. In the present work, we focus on the former point in the context of projected scalar fields, and leave the discussion of covariance matrices, systematics and the entire analysis pipeline to future publications.

Calculating the 3PCF poses a significant challenge, primarily due to its naive scaling as $\mathcal{O}(N^3)$, where $N$ represents the number of objects in the sample, e.g. galaxies in a 3-dimensional catalog or \texttt{HEALPix} pixels \cite{Gorski:2004by} in a weak lensing map. While using k-dimensional structures (kd-trees) or other common algorithms for partitioning data can mitigate computational time, further advancements are needed to compute the 3PCF for several millions of objects. A recent strategy to circumvent this computational bottleneck involves employing a multipole decomposition, which reduces the algorithm's computational scaling time to roughly $\mathcal{O}(N \log N)$, just as the 2PCF scaling. These methods were initially proposed in \cite{Szapudi:2004gg,Zheng:2004eh,Pan:2005ym} for the 3PCF of galaxy counts on spectroscopic catalogs, and further developed in refs.~\cite{Slepian:2015qza,Slepian:2015qwa, Slepian:2016weg,Slepian:2017lpm,Portillo:2017sux,Sunseri:2022cam}. Thereafter, these have been extended to arbitrary {\it N}-point correlations \cite{Philcox:2021bwo,Hou:2021ncj,Philcox:2021eeh}.

The theoretical framework for the 3PCF in galaxy weak lensing has been developed by several researchers over the past two decades \cite{Schneider:2002ze,Zaldarriaga:2002qt,Schneider:2003fy}. These works typically describe the general theory using a brute force approach. 
Consequently, recent progress has been made through the introduction of innovative methodologies aimed at bypassing the need to compute the complete 3PCF. Instead, these approaches focus on constructing summary statistics and estimators to capture specific non-Gaussian features of structure formation. Notable advancements in this direction include the works \cite{Halder:2021itp,Gong:2023nzy,Heydenreich:2022lsa,Linke:2022xnl}. Furthermore, ref.~\cite{Porth:2023dzx} recently applied a similar decomposition in plane waves, as presented in our work, to derive 3-point aperture mass statistics from the KiDS-1000 survey.

The small distortions in the galactic background field caused by a foreground mass distribution, which define the gravitational weak lensing (WL) of galaxies, can be classified into two types: convergence and cosmic shear. Convergence corresponds to isotropic deformations, while cosmic shear corresponds to anisotropic deformations. In this study, we focus on analyzing the 3PCF of scalar fields defined on the sphere, with a specific emphasis on the convergence of WL, even though our methodologies and code are applicable to various types of maps, including galaxy counts and CMB lensing convergence.  In future research, we plan to extend our approach to incorporate the spin-2 cosmic shear. The first part of this work pays attention to the construction of an estimator for the multipoles of the 3PCF in the harmonic basis for full-sky spin-0 maps, together with a description of the first handful of multipoles of the convergence using simulated data. For a complementary ingredient to this first part of the study, we develop a theoretical framework in the harmonic basis to describe the signal of the convergence field based on three different prescriptions: a halo model, standard perturbation theory (SPT) and a model considering the short-range of non-locality \cite{McDonald:2009dh} due to smoothing kernels of density fields (hereafter we call it the EFT model). The EFT model is truly a \textit{primitive} Effective Field Theory description \cite{McDonald:2006mx,McDonald:2009dh,Baumann:2010tm,Vlah:2015sea}, because it does not contain 1-loop corrections to the real space matter power spectrum. Despite this simplification, the EFT model is a notable improvement over SPT, and capable of reproducing results almost as good as the halo model as we show in our results. 
Although the halo model shows a good agreement with statistical estimators from simulated or real data, its primary drawback is the high dependence on various qualitative assumptions, meaning that it cannot, conclusively, offer a comprehensive depiction of clustering. Furthermore, the 2-halo contribution to the matter power spectrum, and also the 3-halo piece in the bispectrum, suffer from similar problems as in the linear power spectrum, since they cannot account for the degradation of the baryon acoustic oscillations, and even the broadband is not well modeled at quasi-linear scales \cite{Bharadwaj:1996qm,Eisenstein:2006nj,Baldauf:2014qfa,Baldauf:2015xfa,DAmico:2022ukl}.  On the other hand, models such as SPT and EFT are based on solid, comprehensive physical theories, yielding theoretical control of the modeling. Indeed, Perturbation Theory (PT)/EFT based models nowadays are widely preferred to analyse the galaxy power spectrum extracted from spectroscopic surveys; e.g. \cite{Ivanov:2019pdj, DAmico:2019fhj,Noriega:2024eyu,Maus:2024sbb}. The situation is certainly more complicated with statistics of projected fields, because integration along the line-of-sight mix the physical scales, and non-linearities become more important, hence it is more common to use halo models in such situations. However, the search for more accurate theoretical models remains ongoing \cite{LSSTDarkEnergyScience:2023qfp}.  That is, in the era of high cosmological precision, we are still in the search of more comprehensive theoretical models of statistics that can prove good accuracy over a wide range of scales. \\
The rest of this work is organized as follows. Section 2 outlines the 3PCF harmonic decomposition for scalar projected fields, while we focus on the weak lensing convergence field in Section 3. In this same section we described the code cBalls, an implementation of this harmonic basis estimator.  We briefly explain in Section 4 how the 3PCF multipoles are obtained from the corresponding bispectrum harmonic coefficients, while in Section 5, we use SPT, EFT and halo descriptions to model the 3PCF coefficients for the convergence field. The comparison of the modeling with synthetic data can be found in Section 6. Finally, we present some remarks and prospects.

\end{section}

\begin{section}{Harmonic decomposition of projected fields}

In this work we focus on a scalar field $X$ defined over the sky. The connected part of its 3PCF is given by
\begin{equation}\label{zetanh}
\zeta(\vt_1,\vt_2;\vnu)=\langle X(\vnu)X(\vnu+\vt_1)X(\vnu+\vt_2) \rangle_c. 
\end{equation}

The position points given by vectors $\vnu$, $\vnu+\vt_1$, and $\vnu + \vt_2$ constitute vertices of triangles, and we compute the average over all such triangles. In the case of statistically homogeneous fields, the 3PCF does not depend on $\vnu$. An unbiased estimator for such correlation is
\begin{equation}\label{zeta}
\hat{\zeta}_\text{hom}(\vt_1,\vt_2)=\int \frac{d^2 \vnu}{A} X (\vnu)X (\vnu+\vt_1)X (\vnu+\vt_2), 
\end{equation}
where $A$ is the area of the considered patch on the sky. 

\revised{About one vertex for each triangle, that we choose to be the ``pivot points'' $\vnu$, we choose an arbitrary unit vector $\vhn$ to define the polar coordinates $(\theta,\phi) = \vt$.} Then, we can construct an estimator for the 3PCF that uses pair-searches instead of triplets by decomposing the fields in a harmonic basis \cite{Szapudi:2004gg,Zheng:2004eh,Philcox:2021eeh},
\begin{equation} \label{Xmdef}
X (\vnu+\vt)=\sum _{m=-\infty}^{\infty}X_m(\theta ; \vnu)e^{im\phi}. 
\end{equation}

\revised{Notice that the tangent planes where the plane-wave bases are defined are different for each pivot on the sphere. Therefore, the}
plane-wave decomposition is a good approximation when the maximum angular scale is just a few degrees at most. In this case, one can approximate the sphere around each pivot $\vnu$ by its tangent plane at that point. This assumption is consistent with the 3PCF convergence signal getting very small at large scales, as we show in this work. If that were not the case, it would be necessary to select an appropriate basis composed of the eigenfunctions of the Laplacian operator on the 2-sphere. Substituting the harmonic decomposition of eq.~\eqref{Xmdef} into the estimator of eq.~\eqref{zeta} we obtain\footnote{After submitting this work for the internal review process at the DESC-LSST collaboration, we discovered the work of \cite{Sunseri:2022cam} where a similar estimator is discussed.}
\begin{align}\label{zetahom}
\hat{\zeta}_\text{hom}(\vt_1,\vt_2)=\int &\frac{d^2\nu}{A}X(\vnu)
\sum_{m_1,m_2}X_{m_1}(\theta_1;\vnu)X_{m_2}(\theta_2;\vnu)   e^{im_1\phi _1}e^{im_2\phi _2}.
\end{align}

We can achieve isotropy by averaging over rotations around the pivot $\vnu$, i.e.,
\begin{equation} \label{zetaisohom}
   \hat{\zeta}_\text{iso, hom}(\theta_{1},\theta_{2},\phi_{1}-\phi_{2}) = \int_0^{2\pi} \frac{d\psi}{2\pi} \,\hat{\zeta}_{hom}(\vt'_{1},\vt'_{2}),
\end{equation}
where the vectors $\vt$ are rotated to $\vt'= R \vt$, which in 2-dimensions only implies translation of the polar angle $\vt=(\theta,\phi)\rightarrow \vt'=(\theta,\phi + \psi)$. 
We now remove the labels $(iso,hom)$ in what follows.
After simple manipulations and using the fact that the $\psi$ integral is not zero only if $m_1=-m_2\equiv m$, the previous expression reduces to
\begin{equation}\label{eq_zeta_harmonic}
\hat{\zeta}(\theta _1,\theta _2;\phi _{12}) = \sum_{m=-\infty}^{\infty} \hat{\zeta}_m(\theta_1,\theta_2) e^{im\phi_{12}},
\end{equation}
where $\phi _{12}\equiv \phi_{1}-\phi_{2}$, and the 3PCF moments or harmonic coefficients are given by
\begin{equation} \label{ubEstimator}
\hat{\zeta}_m(\theta_1,\theta_2)=\int \frac{d^2 \nu}{A}X(\vnu)X_{m}(\theta _1;\vnu)X_{-m}(\theta_2;\vnu).
\end{equation}

While the homogeneous estimator in eq.~\eqref{zetahom} depends on both angles $\phi_1$ and $\phi_2$, the isotropic estimator in eq.\eqref{zetaisohom} depends only on the difference $\phi_1-\phi_2$, which is a key property of these harmonic expansion approaches that allows us to write the $\zeta_m$ multipoles in the simple form of eq.~\eqref{ubEstimator}. Moreover, we consider a single scalar field $X$, but it can be generalized to multiple fields $X^A$, $X^B$ and $X^C$ (where $A$, $B$,... can denote, e.g., different redshift bins, or even different observables) by first choosing a pivot field, e.g. $X^A(\vnu)$, and at the end summing over cyclic permutations of $\{A,B,C\}$ to obtain $\zeta^{ABC}_m$.

Although it is natural to study the full 3PCF (\ref{eq_zeta_harmonic}), in this work we focus on the expansion functions, $\zeta_m$, which can be thought as individual statistics on their own. This has a few benefits. On one hand, the full 3PCF is only recovered in the limit of using an infinite number of multipoles, $m$, as shown in equation (\ref{eq_zeta_harmonic}), but in practice one wants to reduce the complexity of the algorithm by only using a finite and small number of multipoles. The convergence of the sum over $m$ is not, in general, guaranteed. Nonetheless, the components $X_m$ (eq.~\ref{Xmdef}) of the 3PCF multipole decomposition are nothing else that the Fourier transforms of the original data for each radius, so for smooth enough data the Paley-Wiener theorem suggests the asymptotic decay of the modes \cite{PaleyWiener}.   
Actually, for fields such as the weak lensing convergence the series is quickly convergent for most projected triangle configurations, with most of the information being contained in the first handful of multipoles, as will become evident later. However, opting for a description of each multipole separately does not rely on the full 3PCF convergence. On the other hand, these coefficient functions have one less dimension, equivalent in spirit to fixing one distance in the full 3PCF, which then can be easily shown as density plots. To depict all information of the 3PCF one would need a substantial number of multipole plots. However, as it becomes evident later a handful of multipoles is enough to describe most of the meaningful information in our case of study, while still preserving some intuition of the relevant structures in the signal.\footnote{See also \cite{SosaNunez:2020rpe} for further discussions on the 3PCF visualizations.}

Finally, it is possible to model these coefficients separately as we show in section \ref{sec:modeling}. However, before proceeding to this analytic description we would like to employ some time describing a practical estimator for extracting the signal out of data.

\section{The 3PCF signal of the Weak Lensing convergence field}
In order to describe the multipole coefficients (eq.~(\ref{eq_zeta_harmonic})) of the 3PCF for the convergence field in weak lensing, we first introduce the binned estimator and its code implementation. We then describe the signal based on mock catalogues. 

\subsection{3PCF binned estimator}

 For our binned estimator we can proceed similarly as we did in eq.~\eqref{ubEstimator}. In the discrete case with $N$ sample points, or \texttt{HEALPix} pixels, the field $X$ can be written as
\begin{equation}
  X(\vt) = \sum_{i=1}^N w_i x_i \, \dD(\vt-\vt_i), 
\end{equation}
where $x_i$ is the value of the scalar field at the position $\vt_i$. We have considered potential specific weights $\{ w_i \}_{i=1}^N$ that can be added to the sample. At each pivot $\vnu_i$ we construct a radial binning as shown in fig.~\ref{fig:scheme}. Using all points $\vnu_j$ within a radial bin $j$, we obtain the binned moments $m$ of the spin-0 field $X$ around $\nu_i$
\begin{equation}\label{Xm.binned}
X_m(\theta^a;\vnu)=\frac{1}{\sum_{i=1}^{N_p} w_i }\sum _{j=1}^{N_p} w_j x_j \Theta(|\vnu-\vnu_j|;\theta^a)  e^{i m \phi_j} ,
\end{equation}
where $\phi_j$ is the angle with respect to a direction $\vhn$ as in the unbinned case.  The radial binning function $\Theta$ returns 1 when $\vnu_j$ is at a distant $\theta^a$ of $\vnu$ (that is, if $\vnu_j$ lies within the bin $a$) and 0 elsewhere. Notice the vector index $a$ in the variable $\theta$ serves to denote the bin. The $j$-sum runs over the $N_p$ points around the pivot that are assumed to lie inside a maximum radial distance, $\theta_{max}$, not much larger than $200 \,\text{arcmin}$ such that \revised{the flat-sky approximation around $\vnu$ remains accurate. Notice, however, that all pivot points are still moving along the sphere.}
With this notation, the 3PCF multipole $m$ estimator is\footnote{We use matrix notation and omit to write binning indices $a$ and $b$ on $\theta_1^a$ and  $\theta_2^b$.} 
\begin{equation} \label{zetamEst}
  \hat{\zeta}_m(\theta_1,\theta_2) = \frac{1}{\sum_i w_i} \sum_{p=1}^{N} w_j x_j X_m(\theta_1;\vnu_p)  \otimes  X_{-m}(\theta_2;\vnu_p). 
\end{equation}
That is, for each pivot $\nu_p$, we take the exterior product of vectors $X_m$ and $X_{-m}$, multiply by the field evaluated at the pivot, and average over all pivots in the sample.

\begin{figure}
	\begin{center}
	\includegraphics[width=5.5 in]{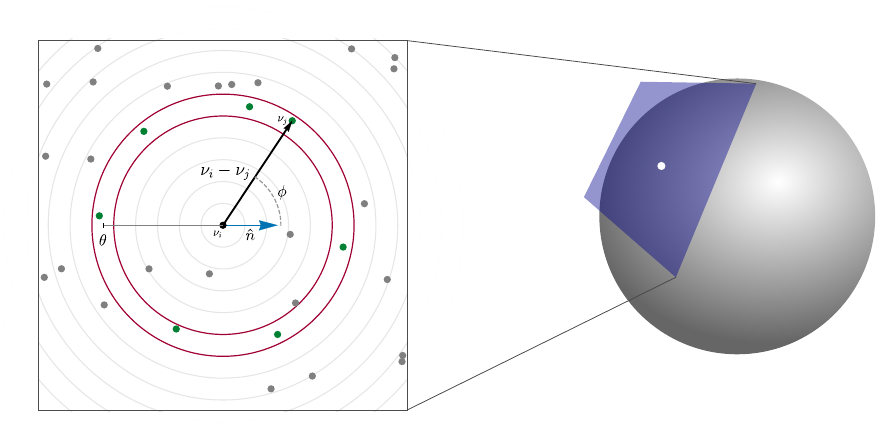}
	\caption{\textit{Binning scheme:} At each pivot point $\vnu_i$, we construct a radial grid. The component $a$ of the vector $X_m(\theta^a;\vnu_i)$ is obtained by evaluating $X$ at each point $\vnu_j$ within bin $a$, multiplying by $e^{i m \phi_j}$, and finally averaging over all the points in that bin. The angles are taken with respect to the unitary vector $\vhn$. \revised{In practice and for the purpose of this work, we use a \texttt{HEALPix} grid, instead of point-like data.}}  \label{fig:scheme}
	\end{center}
\end{figure}

It is convenient to split the $X_m$ coefficients in their real, $Y_m$, and imaginary, $Z_m$, pieces,
\begin{align} \label{Ym.binned}
    Y_m(\theta;\vnu) &= \text{Re}\big[X_m(\theta;\vnu)\big] 
    = \frac{1}{\sum_i w_i }\sum _{j=1}^{N_p} w_j x_j \Theta(|\vnu-\vnu_j|;\theta)  \cos(m \phi_j), \\ \label{Zm.binned}
    Z_m(\theta;\vnu) &= \text{Im}\big[X_m(\theta;\vnu)\big] 
    = \frac{1}{\sum_i w_i }\sum _{j=1}^{N_p} w_j x_j \Theta(|\vnu-\vnu_j|;\theta)  \sin(m \phi_j), 
\end{align}
and write eq.~\eqref{zetamEst} as
\begin{align} \label{zetamEst2}
  \hat{\zeta}_m(\theta_1^a,\theta_2^b) = \frac{1}{\sum_i w_i}  \sum_{p=1}^{N} w_p x_p \Big[ 
  &\big\{ 
  Y_m(\theta_1^a;\vnu_p)Y_m(\theta_2^b;\vnu_p) + Z_m(\theta_1^a;\vnu_p)Z_m(\theta_2^b;\vnu_p) 
  \big\}  \nonumber\\ 
  &+ i \big\{  
  Z_m(\theta_1^a;\vnu_p)Y_m(\theta_2^b;\vnu_p) - Y_m(\theta_1^a;\vnu_p)Z_m(\theta_2^b;\vnu_p) 
  \big\} 
  \Big].
\end{align}

This is the expression we use to obtain the 3-point signal from the data, where we have explicitly included the matrix structure by adding the indices $a$ and $b$ to the binned radial coordinates.  The functions $\cos(m\phi)$ and $\sin(m\phi)$ entering $Y_m$ and $Z_m$ can be efficiently calculated from $\cos(\phi)$ using the Chebyshev polynomials of the first and second kind, respectively. Hence, we never compute trigonometric functions, which are computationally expensive. 
The direction $\vhn$ we use to measure the angles $\phi_j$ is arbitrary, and it necessarily changes when moving from one pivot to another because the space is curved. Hence, the $X_m$ constructed in this way contain a phase which depends on the chosen $\vhn$. However, the combinations $Y_m \otimes Y_m + Z_m \otimes Z_m $ and $Z_m \otimes Y_m - Y_m \otimes Z_m$, as well as the final expression for the 3PCF moments in \eqref{zetamEst} become invariant on the choice of $\vhn$ at each pivot.

In terms of complexity and assuming a single catalog for the three possible fields in the 3PCF, an efficient algorithm would take $\log(N)$ steps to find the $N_\text{p}$ neighbours of each point, with $N$ the total number of pixels or data points in the catalog. Although this number is not constant and depends on each pivot point, it can be roughly approximated by $2\pi\bar{\rho}\theta_{max}^2$. Moreover, once pairs are found around each pivot, there is at least one further calculation for each $m$ to get the harmonic decomposition\footnote{The exact number of calculations at this stage depends on how the code is constructed to calculate the moments in eq.~\eqref{Xm.binned} around each pivot point. In our case we do two separate process calculating the real (eq.~\ref{Ym.binned}) and imaginary (eq.~\ref{Zm.binned}) parts of these moments using Chebyshev coefficients of the first and second kind.} up the maximal number multipole $m_{max}$. The time scaling with the number of multipoles is linear, $t = t_{0} + a m_{max}$, where the time $t_0$ is typically large. Therefore, the resulting algorithm complexity scales as $\mathcal{O}(N\log(N))$ at best, which is the scaling of the standard 2-point algorithm. However, above certain threshold for $m_{max}$ (when $t_0$ becomes negligible in the linear scaling presented above), the time scaling switches to $\mathcal{O}(m_{max} N\log(N))$. Finally, one should notice that as $m_{max}$ approaches $N$, the complexity reduces to that of the 3PCF without the harmonic basis decomposition, $\mathcal{O}(N^2\log(N))$. One way to picture the role of $m_{max}$ in the scaling and what information it brings to the full 3PCF, is to imagine constructing histograms around each pivot point using radial and $n_\phi$ angular bins. Then, to obtain the harmonic coefficients one should take the Fourier transform over these angular bins, resulting in $n_\phi$ operations and $m=n_\phi/2$ final multipoles (remind that one gets coefficients from $-m_{max}$ to $+m_{max}$). This is the same scaling as above. Moreover, in the case of a random sample, one can increase the angular bin resolution so that each bin contains either one or zero data points. In this case, since the Fourier transform does not loose information, these finite number of multipoles carry all the 3PCF information about the sample, with a hierarchy of decaying multipoles as $m$ increases. The fact that the real data distribution in the Universe is mostly driven by a linear theory of very Gaussian initial conditions, suggests that
this convergence argument should approximately hold true, and multipoles with $m>N$ should not provide physical information into the 3PCF. Conversely, one may also use this argument of the Fourier transform to design an algorithm that uses FFTs (which scale as $\log(m)$) to get a more efficient estimator for a large $m_{max}\ll N$.

 Finally, we notice that a mirror reflection (or 2-dimensional parity) of a triangle would change $\phi_{12}\rightarrow -\phi_{12}$, preserving (changing) the sign of the real (imaginary) part of $\zeta_m(\theta_i,\theta_j)$. Breaking isotropy in the 3-dimensions can violate this mirror reflection symmetry in 2-dimensions, which then can be picked up by the imaginary part.  It is the real part of $\zeta_m(\theta_i,\theta_j)$ that contains all the information of isotropic catalogs. Similar ideas have been explored for the CMB. For instance, non-vanishing CMB bispectrum components arising from anisotropic or parity violating 3D fields are characterized by an odd sum of the angular momentum numbers (see for example \cite{Kamionkowski:2010rb}).

\subsection{cBalls code}

We implement the binned-estimator of eq.~\eqref{zetamEst2} into the C-programmed code \texttt{cTreeBalls}\footnote{Publicly available at \href{http://github.com/rodriguezmeza/cTreeBalls.git/}{http://github.com/rodriguezmeza/cTreeBalls.git/}} (\texttt{cBalls}, for short), which computes correlation functions using different tree methods \cite{treeballs}. The codes utilizes \texttt{OpenMP} interface for parallelisation and an octree searching algorithm to find data pairs around the pivot points of the 3PCF triangle configurations. Other tree options, such as kd-trees and ball-trees, are also implemented, as well as averaging methods within the lowest tree cells to reduce computational costs.
The harmonic decomposition is achieved using Chebyshev polynomials of the first and second kind in order to avoid trigonometric calculations, resulting on two polynomial operations per each multipole $m$. The output splits the real and imaginary parts of the 3PCF harmonic coefficients, where the imaginary part can be used to isolate/identify systematics, or other phenomena, that lead to 2-dimensional parity violations, as discussed above. Moreover, the \revised{estimator of eq.~\eqref{zetamEst2} is the same if the pivot points are defined over the sphere or over any other smooth manifold. Therefore, the code can calculate correlations when the pivot points lie over the sphere or on a plane, and using either linear or logarithmic binned-scales for the point separations}. Finally, the package  structure allows for add-ons to perform different pre/post data processing, and extensions to HOC using the harmonic decomposition for non-scalar fields, such as shear, and the use of MPI and GPUs are under development.

In terms of performance, the analysis of a full sky map, with $\sim 200$ million pixels and a maximal searching radius distance of 200 arcmin takes around 40 CPU minutes (wall-clock) on 128 threads of a single Perlmutter-NERSC\footnote{\href{https://docs.nersc.gov/systems/perlmutter/architecture/}{https://docs.nersc.gov/systems/perlmutter/architecture/}.} node. \revised{Assuming that very close\footnote{Formally speaking within a radius of 0.0029 radians or less, which is smaller than the average distance for a catalogue with Nside=4096.} neighboring points to a given pivot share an almost identical list of neighbors allows to further reduce the computational time by an order of magnitude with respect to the original \texttt{cBalls} algorithm, as shown for the curved labeled \texttt{xcBalls} in the left plot of Fig.~\ref{fig:benchmarks}.} A recent,\footnote{Released during the internal DESC-LSST review process for this work.} independent implementation of the 3PCF estimator in the widely used \texttt{TreeCorr}\footnote{Version 5.0, \href{https://rmjarvis.github.io/TreeCorr/}{https://rmjarvis.github.io/TreeCorr/}}  code \cite{treecorr} takes a similar amount of time and the results are compatible with ours.
Actually, for vast amounts of data points the principal CPU time consumption comes from tree searching and summing up the neighbor lists. Furthermore, for each multipole one needs to calculate its corresponding Chebyshev polynomials, which translate into a na\"{\i}ve scaling of $\mathcal{O}(m_{max} N\log(N))$, with $N$ the number of data points and $m_{max}$ the maximal multipole number. In the left panel of figure \ref{fig:benchmarks} we show that the effective power law scaling with the number of data is roughly $N^{1.11}$. This analysis employs Nside \texttt{HEALPix} parameters 256, 512, 1024, 2048 and 4096 with maximum multipole $m_{max}=8$. In the right panel we show the scaling with maximum number of mutipoles, showing a linear scaling $t(m)=1.85 m + 63.7 \,\text{s}$ for $\text{Nside}=1024$. The slope of $t(m)$ is close to $2$, which is the number of operations needed in the recursive relations of the Chebyshev polynomials (first and second kind). To generate both plots we use 128 threads of a single Perlmutter-NERSC node. As a result of this $m$-scaling, the total computational time is 
larger 
than the 2-point functions, since one must obtain the cosine and sine functions through Chebyshev polynomials.  
However, the dependence on $m$ is moderate, being $\sim \! 20 \,\%$ slower to compute up to the multipole $m=8$ than up to $m=1$.

\begin{figure*}
	\begin{center}
	\includegraphics[width=3 in]{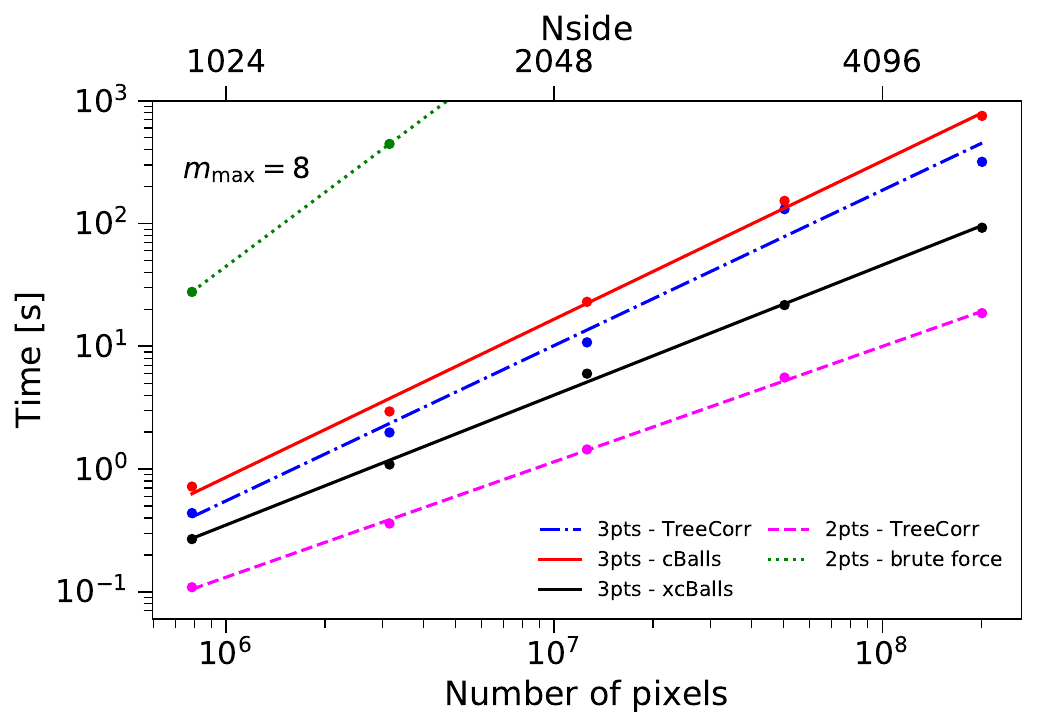}
    \includegraphics[width=3 in]{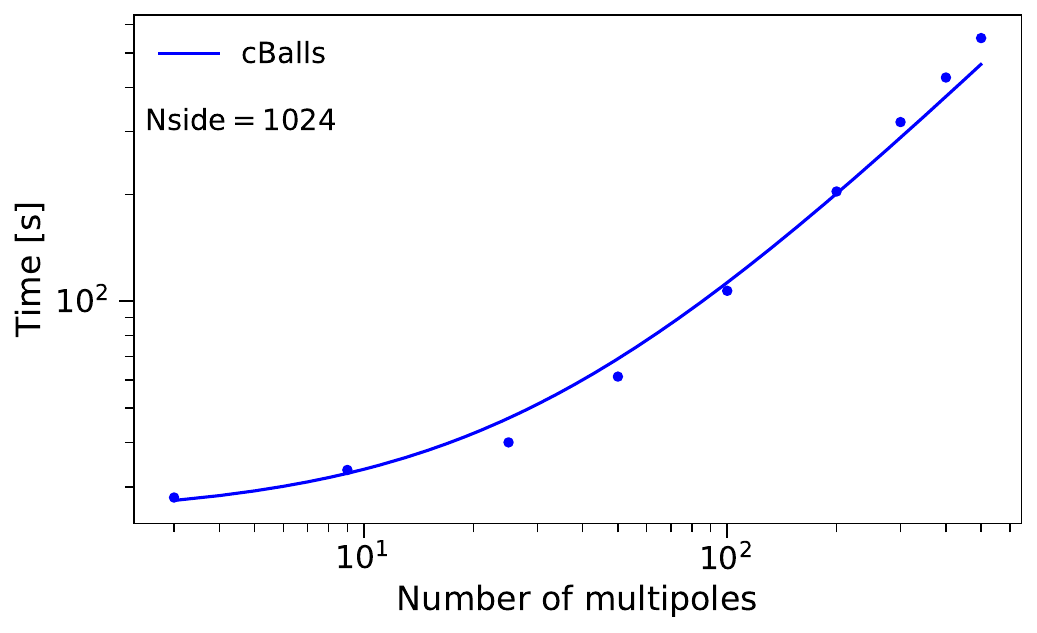}
	\caption{CPU time per thread as a function of the total number of data points or pixels, $N$, for $m_{max}=8$ fixed (left); and as a function of the maximum multipole, $m_{max}$, for Nside=1024 (right). The cBalls code scales as $\sim N^{1.28}$ for octree version, which is closer to the efficient TreeCorr 2pt calculation ($\sim N \log N$) than a brute-force algorithm ($\sim N^2$). \revised{The \texttt{xcBalls} algorithm performs closer to the 2pt calculation ($\sim N^{1.1}$) with a pre-factor that is almost an order of magnitude faster than the original cBalls}. There is growth in the computational time as a function of $m_{max}$ given approximately by $\sim 25+0.87m_{max}$. The benchmarks were obtained using 20 bins for $\theta$, with $\theta_{min}=8$ arcmin, $\theta_{max}=200$ arcmin, in 128 threads on a single CPU node of Perlmutter-NERSC. } \label{fig:benchmarks}
	\end{center}
\end{figure*}

\subsection{T17 Simulations}

This work utilizes the full-sky gravitational lensing simulations by Takahashi et al. (2017) presented in  ref.~\cite{Takahashi:2017hjr}, hereafter referred to as T17.\footnote{Publicly available at \href{http://cosmo.phys.hirosaki-u.ac.jp/takahasi/allsky_raytracing/}{http://cosmo.phys.hirosaki-u.ac.jp/takahasi/allsky\_raytracing/}} These simulations were created using ray-tracing algorithm \cite{Mellier:1998pk,Jain:1999ir} through high-resolution cosmological N-body simulations. They provide various outputs, including convergence and shear maps, for redshifts ranging from $z=0.05$ to $5.3$ at intervals of $150 \hmpc$ comoving radial distance. We specifically employ the data with a \texttt{HEALPix} parameter $\text{Nside}=4096$, corresponding to an angular resolution of 0.86 arcmin and 201,326,592 pixels on a full-sky map. The cosmological model assumed is a flat $\Lambda$CDM with parameters $\Omega_\text{cdm} = 0.233$, $\Omega_\text{b} = 0.046$, $h = 0.7$, $\sigma_8=0.82$ and $n_s = 0.97$. Neutrinos are considered massless. We analyze 108 realizations from the simulations, computing both the mean and error values, that we finally compare against our modelings. We employ three different redshifts: $z=0.5$, 1.0 and 2.0.

\subsection{The $\kappa^3$ multipole signal}

In figure \ref{fig:T17} we display the  real, parity even component of $\hat{\zeta}_m(\theta_1,\theta_2)$, computed using the estimator given by eq.~\eqref{zetamEst2} over the convergence maps of the T17 $z=0.5$ redshift bin. For enhanced clarity in visualization we have set $\theta_2 = 61 \, \text{arcmin}$. This plot shows the mean of the 108 realizations for $\zeta_m$ up to multipole $m=8$, while the vertical error bars denote the standard deviations of the realizations.  These plots exemplify typical outcomes, showcasing characteristic patterns of power-law scalings away from the diagonal ($\theta_1=\theta_2$, i.e. isosceles triangles). If one fixes one of the scales, e.g. $\theta_2$ as in figure \ref{fig:T17}, then there is one power law behaviour, which is a straight-line in a log-log plot, for scales $\theta_1<\theta_2$ and a different slope for $\theta_1>\theta_2$. The behaviour near the diagonal is also enhanced by squeezed configurations which only happen for isosceles or close to isosceles triangles.
It is important to stress that our code overcounts the signal along this diagonal, since when doing the product of the two $X_m$ in eq.~(\ref{zetamEst}) it does not remove the information of a point with itself \footnote{The Treecorr adaptation of a similar estimator cleverly subtracts these extra counts when $\theta_1=\theta_2$. See the new release \url{https://rmjarvis.github.io/TreeCorr/_build/html/changes.html}.}. However, as the data sample grows this overcounting is less significant due to the larger number of pairs around each pivot, which for the results of figure \ref{fig:T17} does not account for more than one percent overestimate of the peak height. Further, the monopole $\zeta_0$ is always the dominant coefficient that, except at the peak position, accounts for around half of the total signal or more. After the monopole, the dominant component is the quadrupole $\zeta_2$, followed by the dipole $\zeta_1$. Thereafter, the rest of the multipoles decay in amplitude with $m$. One may wonder if a larger quadrupole is part related to the spin-2 nature of gravity.

\begin{figure}
	\begin{center}
    \includegraphics[width=4 in]{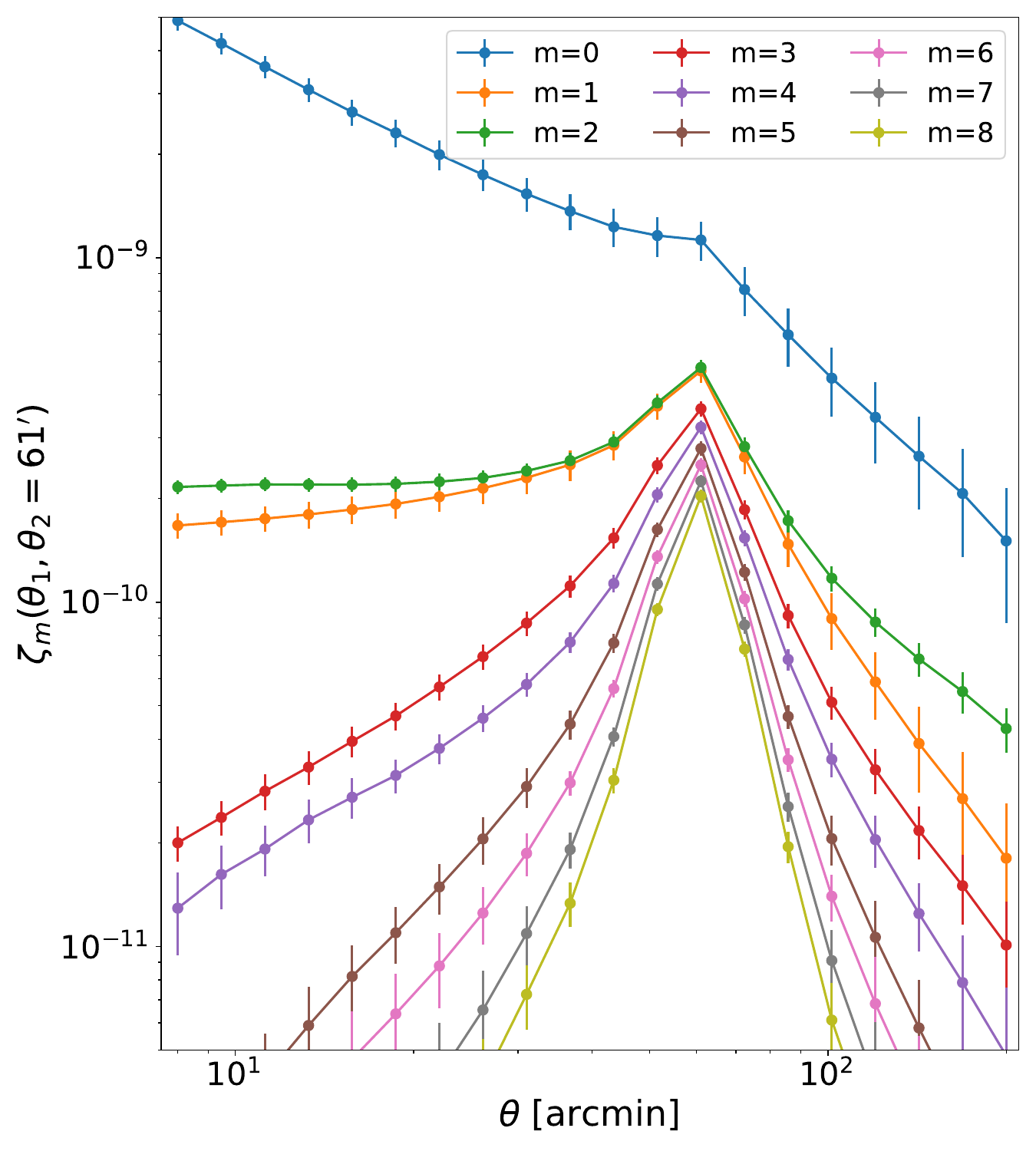}
	\caption{Real part of multipoles $\zeta_{m}(\theta_1,\theta_2)$ at fixed $\theta_2=61\,\text{arcmin}$ for multipoles $m=1,\dots,8$. We use T17 simulations with \texttt{HEALPix} parameter $\text{Nside}=4096$ at redshift $z=0.5$. We display the mean and standard deviations of the 108 realizations available. The monopole has the largest amplitude and higher multipoles are ordered hierarchically, except for the quadrupole which is above the dipole.}  \label{fig:T17}
	\end{center}
\end{figure}

\begin{figure}
	\includegraphics[width=7 in]{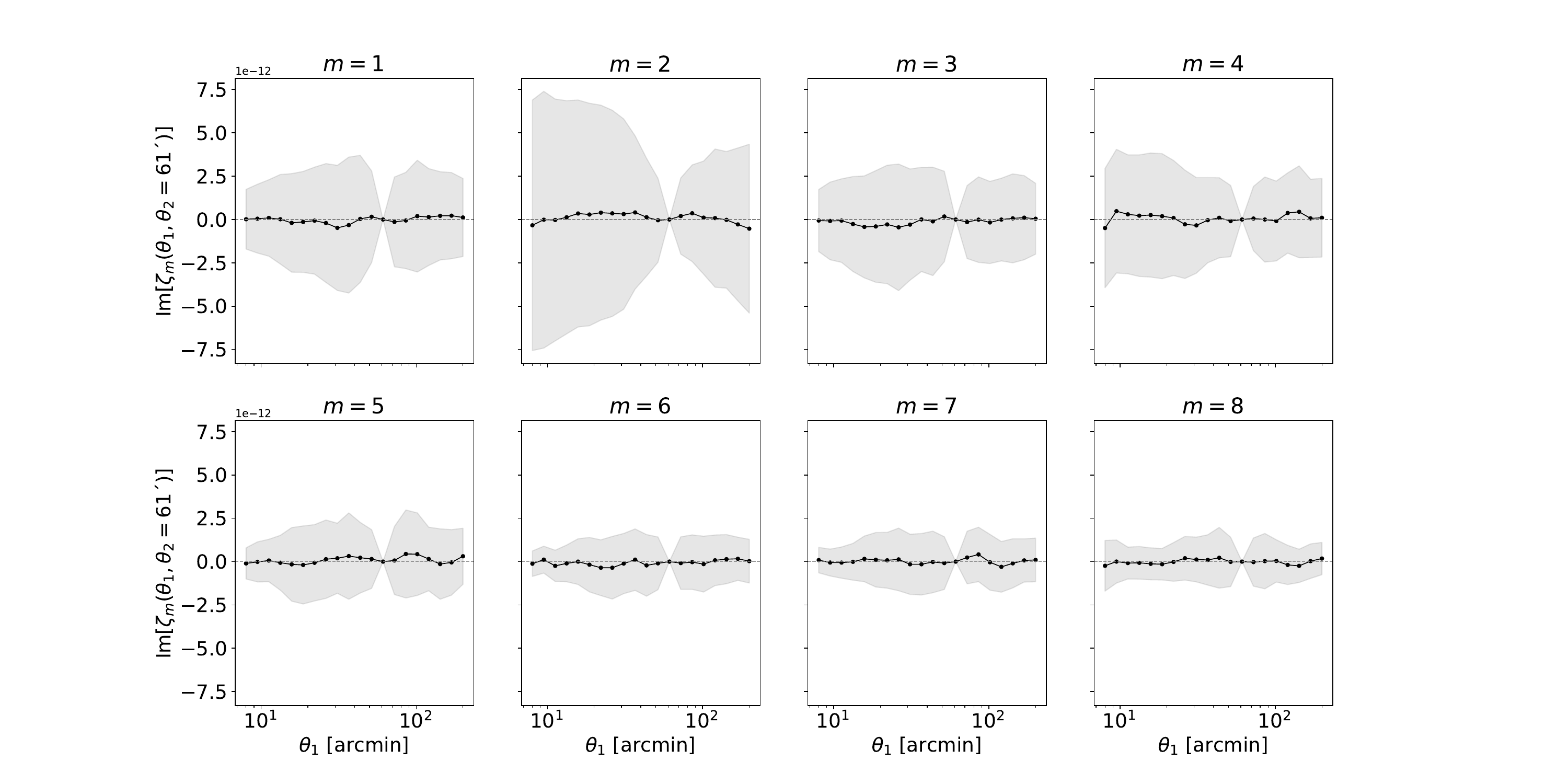}
		
	\caption{Imaginary part of multipoles $\zeta_{m}(\theta_1,\theta_2)$ at fixed $\theta_2=61\,\text{arcmin}$ for multipoles $m=1,\dots,8$. We use the 108 realizations in T17 using the \texttt{HEALPix} parameter $\text{Nside}=4096$ at redshift $z=0.5$. The means are shown by the black lines, and the gray shadows represent the standard deviation about the mean. The imaginary part $\text{Im}[\zeta_{m}]$ is always zero at the diagonal $\theta_1=\theta_2$.}  \label{fig:T17impart}
\end{figure}

In fig.~\ref{fig:T17impart} we show the imaginary part for some of the coefficients $\zeta_m$. That is, we take the second term in eq.~\eqref{zetamEst2}, and display the 1-dimensional plots of $\text{Im}\big[\zeta_m(\theta_1, \theta_2=61')\big]$ for the first eight multipoles. All the correlations are consistent with zero, which is a signature of data obeying parity on the projected sphere, as expected from isotropic simulations. At the diagonal points $\theta_1=\theta_2$ the functions  $\text{Im}\big[\zeta_m\big]$ are exactly zero.

\end{section}

\begin{section}{Analytic modeling of the 3PCF}

Let us consider the theoretical moments $\zeta_m$, defined through the multipole expansion
\begin{align}
    \zeta(\theta_{1},\theta_{2},\phi_{1}-\phi_{2}) &= \sum_{m=-\infty}^{\infty} \zeta_{m}(\theta_{1},\theta_{2})e^{im(\phi_{1}-\phi_{2})}, 
\end{align}
with
\begin{equation}
\zeta_m(\theta_1,\theta_2) = \int_{0}^{2\pi} \, \frac{d\phi}{2\pi} \zeta(\theta_1,\theta_2,\phi) e^{- i m \phi},
\end{equation}
where we used the orthogonality conditions of the plane waves,
$ \int_0^{2\pi} d\phi \, e^{i m \phi} e^{i n \phi} = 2 \pi \delta_{m,-n}$.

Conversely, we can work in Fourier space. The 2-dimensional (or projected) bispectrum $B(\vell_1,\vell_2,\vell_3)= B(\ell_1,\ell_2,\varphi)$,\footnote{We adopt the Fourier transform conventions  
$    f(\vell) = \int d^2\theta \, f(\vt) e^{-i \vt \cdot \vell}$ and 
 $   f(\vt) = \int \frac{d^2 \ell}{(2 \pi)^2} f(\vell) e^{i \vt \cdot \vell}$.}
which as in the 3PCF case can be characterized by two of their sides, $\ell_1$ and  $\ell_2$, and the opening angle between these two sides, $\varphi \equiv \varphi_1 - \varphi_2 = \cos^{-1}(\hat{\ell}_1\cdot\hat{\ell}_2)$. This bispectrum is again expanded in plane waves
\begin{align}\label{Bexp}
B(\ell_1,\ell_2,\varphi) &= \sum_{m=-\infty}^{\infty} B_m(\ell_1,\ell_2) e^{i m \varphi}. 
\end{align}

Using the plane wave expansion in cylindrical coordinates (polar coordinates in 2-dimensions),
\begin{equation} \label{PWexp}
 e^{i \vell \cdot \vt} = \sum_{m=-\infty}^\infty i^m J_m( \ell \theta ) e^{i m \beta}, \quad \text{with} \quad \cos{\beta} = \hat{\ell} \cdot \hat{\theta}, 
\end{equation}
and the property of the Bessel functions $J_{-m} (x) = (-1)^m J_m(x)$, 
is straightforward to arrive to a relation between multipoles $\zeta_m$ and $B_m$, namely
\begin{align} \label{BnToZn}
    \zeta_{m}(\theta_{1},\theta_{2}) &= (-1)^{m} \int \frac{\ell_{1}d\ell_{1}\ell_{2}d\ell_{2}}{(2\pi)^{2}} J_{m} (\ell_{1} \theta_{1} ) J_{m}( \ell_{2} \theta_{2} )  B_{m}(\ell_1,\ell_2),
\end{align}
which is the 2-dimensional equivalent to the formula found in \cite{Szapudi:2004gg} for  scalar fields defined over 3-dimension. Equation \eqref{BnToZn} is a double Hankel transform that is challenging to efficiently perform with enough precision due to its oscillatory nature. In the next section we refer to two different methods we employ to tackle this integral.

For a real scalar field $X$, it follows that $\zeta_m = \zeta_{-m}^*$. However, $\zeta_m = \zeta_{-m}$ is not primarily guaranteed. To ensure this equality, one must assume parity, specifically $\zeta(\theta_1,\theta_2,\phi) = \zeta(\theta_1,\theta_2,-\phi)$. Therefore, any observation of $\zeta_m \neq \zeta_{-m}$ would indicate either a violation of parity in 2-dimensions or the presence of a systematic error not considered in the error analysis. As we mentioned in the previous sections, a potential violation of parity in two dimensions may not be fundamental, but can be due to a departure from isotropy in three dimensions. 
In the following we will consider primarly the even parity piece of the 3PCF. In such a case 
the multipoles of the 3PCF are real, and we can rewrite 
\begin{align}
    \zeta(\theta_{1},\theta_{2},\phi_{12}) &= \sum_{m=-\infty}^{\infty} \zeta_{m}(\theta_{1},\theta_{2})\cos(m \phi_{12}) \\
    &= \sum_{m=-\infty}^{\infty} \zeta_{m}(\theta_{1},\theta_{2})[\cos(m\phi_{1})\cos(m\phi_{2})  
     + \sin(m\phi_{1})\sin(m\phi_{2})],
\end{align}
which is the predicted signal found in the real part of the measurement, eq.~\eqref{zetamEst2}.
On the other hand, the odd parity piece of the 3PCF is 
\begin{align}
    \zeta^\text{parity-odd}(\theta_{1},\theta_{2},\phi_{12}) &=  \frac{1}{2}\big(\zeta(\theta_1,\theta_2,\phi_{12})-\zeta_m(\theta_1,\theta_2,-\phi_{12})\big) \\
    &= i \sum_{m=-\infty}^{\infty} \zeta_{m}(\theta_{1},\theta_{2})\sin(m \phi_{12}) \nonumber\\
    &= i \sum_{m=-\infty}^{\infty} \zeta_{m}(\theta_{1},\theta_{2})[\sin(m\phi_{1})\cos(m\phi_{2})   - \cos(m\phi_{1})\sin(m\phi_{2})],
\end{align}
which corresponds to the imaginary part of eq.~\eqref{zetamEst2}.

To conclude the analytical 3PCF, we require a theoretical framework for the bispectrum. In the following section, we explore various methods to achieve this.

\end{section}

\begin{section}{Modeling the Weak Lensing convergence}\label{sec:modeling}

The convergence observed in weak lensing of galaxies arises from the distortion of background galaxy shapes induced by the gravitational pull of foreground mass. This phenomenon quantifies the degree to which light paths deviate due to the distribution of mass along the line of sight. It is easily understood as the projection over the sky of the matter density field lying in our past light cone,
\begin{equation}
    \kappa(\vt) = \int_0^\infty d\chi \, q(\chi) \delta_m(\chi \theta; \chi),
\end{equation} 
with $\delta_m$ the matter density fluctuation, $\chi$ the comoving distance that we use hereafter as a time coordinate. For large $\chi$ the function $q(\chi)$ tends to zero, hence the upper limit of the integral is not relevant, which for a finite galaxy sample can be taken up to its limiting comoving distance. The lens efficiency is
\begin{equation}
 \revised{  q(\chi) = \frac{3 \Omega_m H^2_0}{2c^2}  \frac{\chi}{a(\chi)}\int_\chi^\infty d\chi' W_g(\chi') \frac{\chi'-\chi}{\chi'}, }
\end{equation}
where $\Omega_m$ is the matter abundance today, $a$ the scale factor, $H_0$ the Hubble constant and $W_g(\chi)$ is the photo-z distribution of the number of galaxies.

In this section we analytically model the weak lensing convergence 3PCF. The starting point is the 3-dimensional matter bispectrum $B_\delta^\text{3D}(\vk_1,\vk_2,
\vk_3)$. Using the Limber approximation \cite{1953ApJ...117..134L,Kaiser:1991qi} one can compute the convergence bispectrum by integrating along the line-of-sight
\begin{align} \label{Bkappa}
B^\kappa(\vell_1,\vell_2,\vell_3) &= \int d\chi \frac{q^3(\chi)}{\chi^4} B_\delta^\text{3D} \left(\frac{\vell_1}{\chi},\frac{\vell_2}{\chi},\frac{\vell_3}{\chi};\chi \right).
\end{align}

From here, we can take different routes to compute the matter bispectrum. In this work we use STP, EFT and a halo model, as presented in the following subsections.

\begin{subsection}{SPT and EFT}

The tree-level matter bispectrum is
\begin{align} \label{Bm3d}
B_\delta^\text{3D}(\vk_1,\vk_2,\vk_3;\chi)  &= 2 F_2(\vk_1,\vk_2) P_L(k_1;\chi)P_L(k_2;\chi) + \text{cyclic},
\end{align}
where ``cyclic'' denotes the sum over cyclic permutations of wave-vectors ($(\vk_1,\vk_2) \rightarrow (\vk_2,\vk_3) \rightarrow (\vk_3,\vk_1)$), 
with $P_L$ the linear matter power spectrum and $F_2$ the second order SPT kernel \cite{Bernardeau:2001qr},
\begin{equation} \label{F2Kernel}
      F_2(\vk_1,\vk_2) =  \frac{5}{7} + \frac{2}{7}\frac{(\vk_1\cdot\vk_2)^2}{k_1^2 k_2^2} + \frac{\vk_1\cdot \vk_2}{2 k_1 k_2}\left( \frac{k_1}{k_2} + \frac{k_2}{k_1} \right). 
\end{equation}
This expression can be further simplified by noting that it only depends on the cosine of the angle $\varphi$ between vectors $\vk_1$ and $\vk_2$, and the ratio of their norms $k_1/k_2$. Hence, evaluating at wave vectors  $\vk_{1,2} = \vell_{1,2}/\chi$, we obtain
\begin{align} \label{F2ell}
F_2\left(\frac{\vell_1}{\chi},\frac{\vell_2}{\chi}\right) &= \frac{6}{7} +  \frac{1}{4} \left( \frac{\ell_1}{\ell_2}+\frac{\ell_2}{\ell_1}\right) (e^{i \varphi}   +    e^{-i \varphi}) +  \frac{1}{14} (e^{i 2 \varphi}   +    e^{-i 2 \varphi} ) \\
&= F_2\left(\vell_1,\vell_2\right).
\end{align}

Notice that Einstein-de Sitter (EdS) perturbative kernels $F_n(\vk_1,\dots,\vk_n)$ are invariant against an overall scale tranformation, $\vk_{1,\dots,n} \rightarrow  c \,\vk_{1,\dots,n}$, and hence they remain time-independent when one writes $\vk_{1,\dots,n}=\vell_{1,\dots,n}/\chi$ to perform projections along the line-of-sight. For $\Lambda$CDM kernels, instead, the $F_2$ kernel does depend on $\chi$ because the factors $5/7$ and $2/7$ in eq.~\eqref{F2Kernel} become multiplied by time-dependent functions \cite{Bernardeau:2001qr}. Moreover, in the presence of additional scales, as in those introduced by some modified gravity models or in the presence of massive neutrinos, these functions become also scale-dependent \cite{Aviles:2017aor,Aviles:2020cax,Aviles:2023fqx}. However, when EdS is a good approximation one can safely use eq.~\eqref{F2ell}, and in such a case the kernel $F_2$ can be pulled out of the $\chi$ integral in eq.~\eqref{Bkappa}. This is a key property in SPT that will permit us to decompose the 2-dimensional Hankel transform into the multiplication of two single Hankel transforms.

The bispectrum momenta in eq.~\eqref{Bexp}, before performing the cyclic permutations (that is, considering only the first term in the rhs of eq.~\eqref{Bm3d}) become
\begin{align}
B_0^{pc}(\ell_1/\chi,\ell_2/\chi) &= \frac{12}{7} P_L(\ell_1/\chi )P_L(\ell_2/\chi ), \\
B_{1}^{pc}(\ell_1,\ell_2) &= B_{-1}(\ell_1,\ell_2) = \frac{1}{2} \left( \frac{\ell_1}{\ell_2}+\frac{\ell_2}{\ell_1}\right) P_L(\ell_1/\chi )P_L(\ell_2/\chi ),   \\ 
B_2^{pc}(\ell_1/\chi,\ell_2/\chi)&=B_{-2}(\ell_1/\chi,\ell_2/\chi) = \frac{1}{7} P_L(\ell_1/\chi )P_L(\ell_2/\chi ), \\
B_{|m|>2}^{pc}(\ell_1/\chi,\ell_2/\chi) &= 0, 
\end{align}
where $pc$ stands for the precyclic momenta. Using the Hankel transformation \eqref{BnToZn}, we obtain

\begin{align}
\zeta_0^{pc}(\theta_1,\theta_2) &= \frac{12}{7}   A_{0}(\theta_1,\theta_2),\\
\zeta_1^{pc}(\theta_1,\theta_2) =\zeta_{-1}^{pc}(\theta_1,\theta_2) &= - \frac{1}{2}A_{1}(\theta_1,\theta_2), \\
\zeta_2^{pc}(\theta_1,\theta_2) =\zeta_{-2}^{pc}(\theta_1,\theta_2) &= \frac{1}{7} A_{2}(\theta_1,\theta_2), \\
\zeta_{|m|>2}^{pc}(\theta_1,\theta_2) &=0.
\end{align}
The $A_{n}$ functions are defined by
\begin{align} 
  A_{0}(\theta_1,\theta_2) 
  & =   \int d\chi \frac{q^3(\chi)}{\chi^4} \Xi^{[0]}(\theta_1,\chi)\Xi^{[0]}(\theta_2,\chi), \label{X0}\\
  A_{1}(\theta_1,\theta_2) 
  & = \int d\chi \frac{q^3(\chi)}{\chi^4} \Xi^{[1,1]}(\theta_1,\chi)\Xi^{[1,-1]}(\theta_2,\chi)  + (\theta_1 \leftrightarrow \theta_2), \label{X1} \\
  A_{2}(\theta_1,\theta_2) 
  & =  \int d\chi \frac{q^3(\chi)}{\chi^4} \Xi^{[2]}(\theta_1,\chi)\Xi^{[2]}(\theta_2,\chi),\label{X2} 
\end{align}
where one has to perform the line-of-sight integrals of the generalized correlation functions
\begin{align}
\Xi^{[n,m]}(\theta,\chi) &= \int \frac{\ell d\ell}{2 \pi}\ell^{m} P_L(\ell/\chi) J_n(\theta \ell), \label{Xinm}\\ \qquad  \Xi^{[n]}(\theta,\chi) &= \Xi^{[n,0]}(\theta,\chi)\label{Xin0}.
\end{align}

The structure of the kernel $F_2$ has allowed us to factorize the 2-dimensional integral \eqref{BnToZn} into two 1-dimensional Hankel transformations, significantly simplifying the numerical calculations. These integrals can be computed using standard FFTLog methods \cite{TALMAN197835,Hamilton:1999uv}.

\begin{figure*}
	\begin{center}
	\includegraphics[width=6 in]{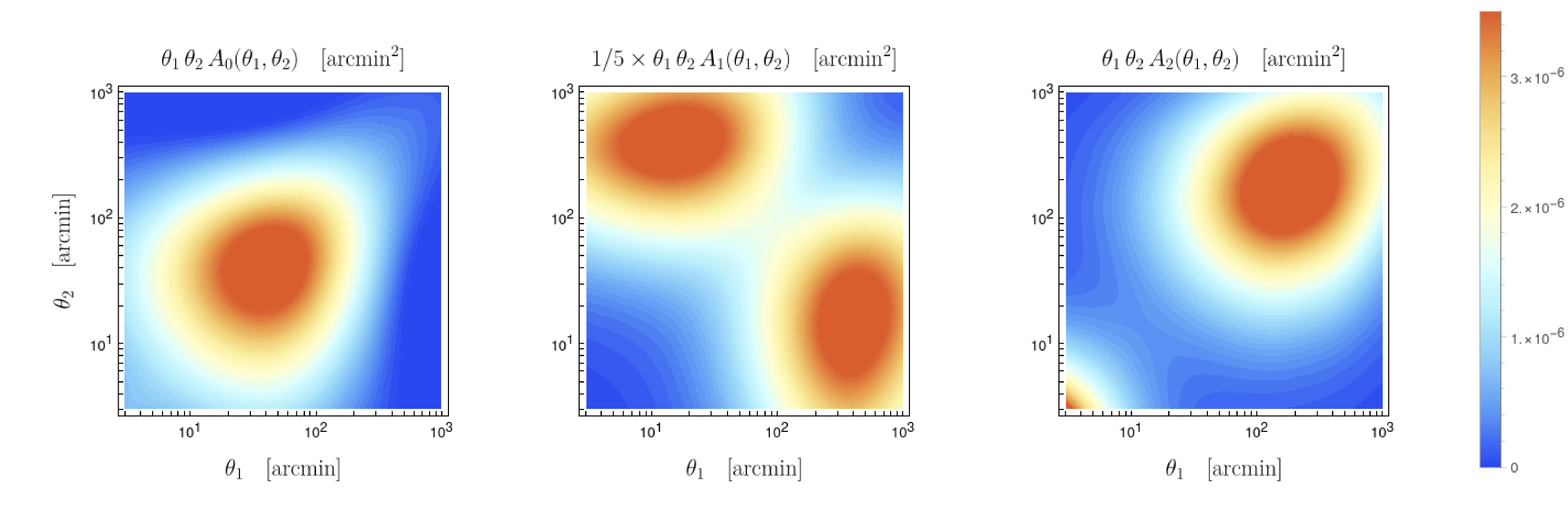}
 \caption{	Functions $A_m(\theta_1,\theta_2)$ computed using the method presented in appendix \ref{sect:FFTLOG3pts}.} \label{fig:Xm}
	\end{center}
\end{figure*}

Then, the pre-cyclic (pc) 3PCF can be written as
\begin{equation} \label{zetapc}
    \zeta_\text{pc}(\theta_1,\theta_2, \phi)= \sum_{m=-2}^2 a_m A_m(\theta_1,\theta_2) e^{i m \phi},
\end{equation}
with $a_0=12/7$, $a_{1}=a_{-1}=-1/2$, $a_{2}=a_{-2}=1/7$.

To perform the line-of-sight integrals in eqs.~\eqref{X0}-\eqref{X2} together with eq.~\eqref{Xinm}, we use an FFT method tailored for this work and presented in appendix \ref{sect:FFTLOG3pts}. The functions $A_n$ are the mathematical building blocks of the pre-cyclic correlation function, resulting in a mixing of  scales in the complete correlation function. In figure \ref{fig:Xm}  we plot the $A_n$ functions using the FFTLog parameters $N=256$, $k_\text{min}=10^{-6}$, $k_\text{max}=100$, with bias $\nu_b=-1.3$ for function $A_1$, and $\nu_b=-0.3$ for $A_0$ and $A_2$, with a range which is beyond the flat-space approximation for the binned-estimator (\ref{Xm.binned}) in order to show the full structure of these building-block functions.  

The total 3PCF is obtained by cyclic summing the pre-cyclic 3PCF
\begin{align}
\zeta(\theta_1,\theta_2,\phi_{12}) &= \sum_{m=-\infty}^\infty  \big[\zeta^\text{pc}_{m}(\theta_1,\theta_2) e^{i m \phi_{12}} + \zeta^\text{pc}_{m}(\theta_2,\theta_3) e^{i m \phi_{23}}  + \zeta^\text{pc}_{m}(\theta_3,\theta_1) e^{i m \phi_{31}}  \big], \label{postc3pcf1}
\end{align}
where $\theta_3$, $\phi_{23}$ and $\phi_{31}$ can be written as functions of $\theta_1$, $\theta_2$ and $\phi_{12}\equiv \phi$. \revised{We want to write the final expression for the 3PCF as}
\begin{align}
\revised{ \zeta(\theta_1,\theta_2,\phi_{12})    = \sum_{m=-\infty}^\infty  \zeta_{m}(\theta_1,\theta_2) e^{i m \phi_{12}}.}\label{postc3pcf2} 
\end{align}
\revised{Hence, we have to express the last two terms  in Eq.\eqref{postc3pcf1} using our original base of plane waves, $e^{i m \phi_{12}}$.}
%
%
%
\revised{To do this, } we define the projections of pre-cyclic multipoles $M$ onto multipole $m$
\begin{align}
 \mI^{(m,M)}(\theta_1,\theta_2) &\equiv \frac{1}{2 \pi} \int_{0}^{2\pi} d\phi_{12} \Big[ \zeta^{pc}_M(\theta_2,\theta_3)e^{i M \phi_{23}} 
 + \zeta^{pc}_M(\theta_3,\theta_1)e^{i M \phi_{31}} \Big] e^{-i m \phi_{12}}. 
\end{align}

Then,
\begin{align} \label{pstcyc3PCF}
\zeta_m(\theta_1,\theta_2) &= \zeta^{pc}_{m}(\theta_1,\theta_2) + \sum_{M=-2}^{2}  \mI^{(m,M)}(\theta_1,\theta_2). 
\end{align}

For parity even statistics 
we obtain the momenta
%
\begin{align}  
\zeta_m(\theta_1,\theta_2)  &=  \zeta^{pc}_{m}(\theta_1,\theta_2) +  \sum_{M=0}^{\infty} \frac{2-\delta_{M,0}}{2} \mJ^{(m,M)}(\theta_1,\theta_2),
\end{align}
with
\begin{align}
\mJ^{(m,M)}(\theta_1,\theta_2) &=  \frac{1}{\pi} \int_{0}^{2\pi} d\phi_{12} \Big[ \zeta^{pc}_M(\theta_2,\theta_3) \cos(M \phi_{23} ) + \zeta^{pc}_M(\theta_3,\theta_1)\cos(M \phi_{31} ) \Big] \cos( m \phi_{12}).
\end{align}
This is the final equation we use to compute the multipoles $\zeta_m$ using SPT.

A crucial theoretical aspect beyond perturbation theory lies in EFT \cite{McDonald:2006mx,McDonald:2009dh,Baumann:2010tm}. Traditionally, it emerges as a necessity due to the removal of the cutoff scale in loop integral regularization, which demands the addition of counterterms with appropriate functional forms. This approach effectively captures the influence of small-scale fluctuations on large-scale structures, profoundly impacting the modeling of observed power spectra and other statistics. 
However, an equivalent method of incorporating EFT contributions, is to notice that even for the dark matter field there is a short-range of non-locality \cite{McDonald:2009dh} that arise from the smoothing of the density field, even for the pure dark matter case. This smoothing should exist not only for technical reasons, but also it is needed theoretically because the fluid evolution equations are not valid below certain scales at which the Boltzmann hierarchy cannot be truncated at the second moment. The effect of this smoothing is to drop out the small scales of the theory at the field level. 
Consequently, a strategy for incorporating EFT (or short-range of non-locality) corrections even at the tree-level statistics involves substituting the dark matter overdensities by their smooth versions in the limit where these are small compared to the scales of interest: $\delta_\text{cdm}(\vx) \rightarrow \delta_{\text{cdm,smooth}}(\vx) = \delta_\text{cdm}(\vx) + \alpha^\text{EFT}(t) \nabla^2 \delta_\text{cdm}(\vx)$, with the parameter $\alpha^\text{EFT}(t)$ treated as a free time-dependent function of the theory, that ultimately models the backreaction effects of the small scales, removed from the theory, over the large scales. 
In our above derived equations, this can be accounted by replacing the linear power spectrum by
\begin{equation}
    P_L(k,z) \rightarrow \big(1 + \alpha^\text{EFT}(z) k^2 \big) P_L(k,z).
\end{equation}
We do this to construct a (primitive) EFT model for the 3PCF multipoles by performing this replacement into eq.~\eqref{Bm3d}. More comprehensive models for the matter bispectrum can be found in the EFT literature, e.g.  \cite{Baldauf:2014qfa,DAmico:2022ukl}.

\end{subsection}

\begin{subsection}{Halo model}

\begin{figure*}
	\begin{center}
	\includegraphics[width=6 in]{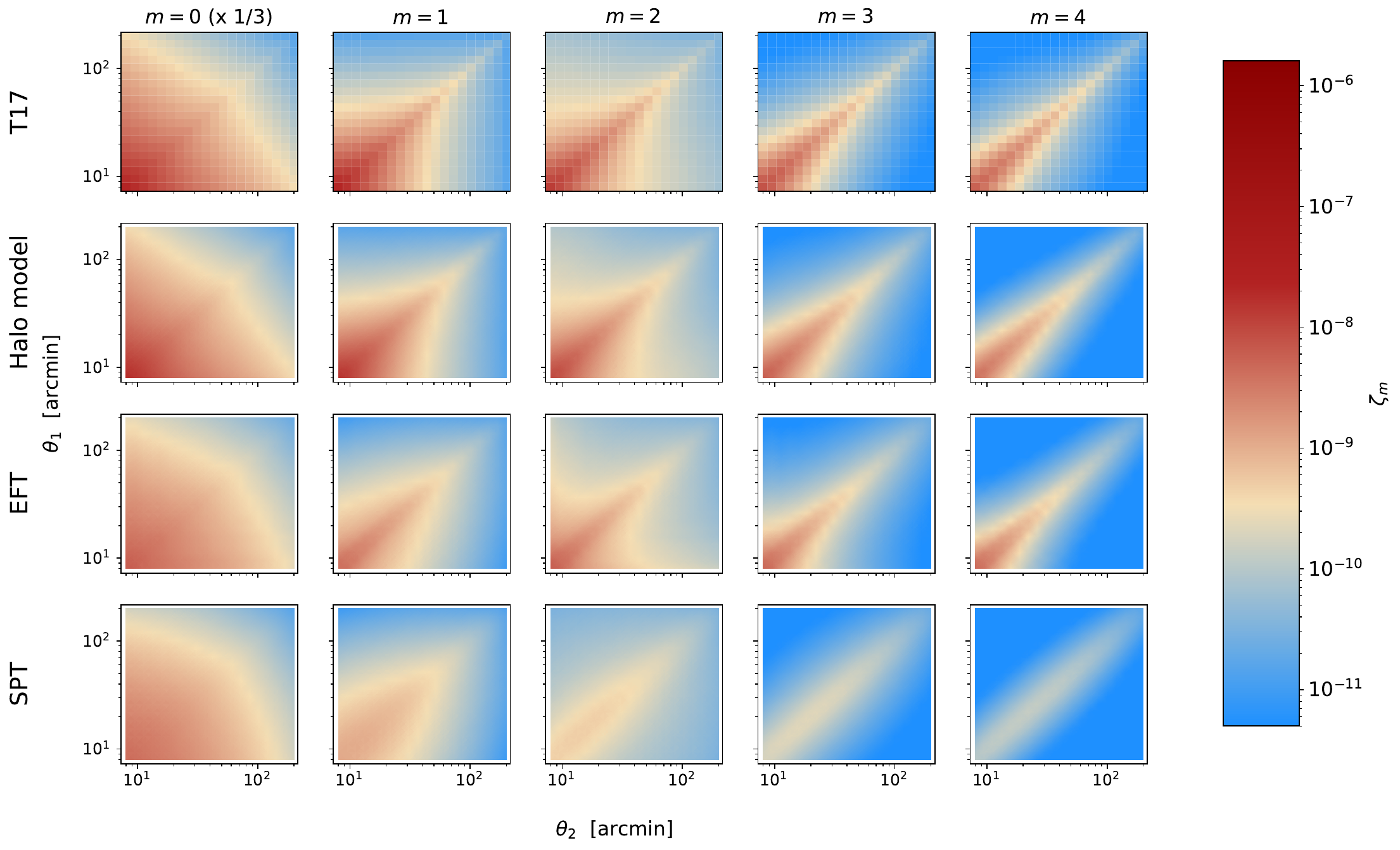}
	\caption{Multipoles $\zeta_{m=0,1,2,3}(\theta_1,\theta_2)$ of the 3PCF obtained from the simulations T17, and using the Halo model, EFT and SPT for redshift bin centred at $z=0.5$.} \label{fig:squareszs9}
	\end{center}
\end{figure*} 

In WL, it is common practice to use halo models \cite{Seljak:2000gq,Ma:2000ik} to describe the matter density field. The power spectrum using these halo models can then be integrated along the line-of-sight to obtain the convergence and shear statistics. As a consequence, these models mix scales across a wide range, especially at high wave numbers, $k \gtrsim 0.5$, where they outperform theoretical approaches like EFT. 
 \revised{However, despite the halo models offer better accuracy, they are not comprehensive theories derived from first principles.}

To compute the matter bispectrum $B_\delta^\text{3D}(\vk_1,\vk_2,\vk_3)$ we use the \texttt{BiHalofit} code, released in \cite{Takahashi:2019hth}\footnote{\href{http://cosmo.phys.hirosaki-u.ac.jp/takahasi/codes_e.htm}{http://cosmo.phys.hirosaki-u.ac.jp/takahasi/codes\_e.htm}}, which even though is not based on the halo model it has a similar rescaling of the linear bispectrum. This fitting formula was calibrated using cosmological N-body simulations of 41 $w$CDM models cosmologies around the Planck 2015 best-fit parameters. In the Planck model, the formula is accurate to $10\,\%$ up to $k=10 \,\hmpci$. While for the dark energy models with constant equation of state, the accuracy is reduced to $20\,\%$.

We assume isotropy and homogeneity such that the bispectrum only depends on the absolute values of the wavenumbers, or alternatively on two the size of two sides $k_1$ and $k_2$ and the angle between them $\varphi$. We use eqs.~\eqref{Bkappa} and \eqref{Bexp} to calculate the convergence bispectrum multipoles as
\begin{align}
    B^\kappa_m(\ell_1,\ell_2) &= \int_0^{\chi_\text{max}} d \chi \frac{q^3(\chi)}{\chi^4}  \int_0^{2\pi} \frac{d\varphi}{2 \pi} B_\delta^{3d}\left(\frac{\ell_1}{\chi},\frac{\ell_2}{\chi},\varphi;\chi \right) e^{-i m \varphi},
\end{align}
where the interval $(0,\chi_\text{max})$ covers the  support of the function $q(\chi)$. 

To obtain the multipoles $\zeta_m^\kappa(\theta_1,\theta_2)$, we need to perform the 2-dimensional Hankel transformations as described in equation \eqref{BnToZn}. This involves utilizing the 2D-FFTLog method outlined in \cite{Fang:2020vhc}, which provides a C code for carrying out these transformations.\footnote{\href{https://github.com/xfangcosmo/2DFFTLog}{https://github.com/xfangcosmo/2DFFTLog}}

\end{subsection}

\end{section}

\begin{section}{Modeling accuracy}

\begin{figure*}
	\begin{center}
	\includegraphics[width=6 in]{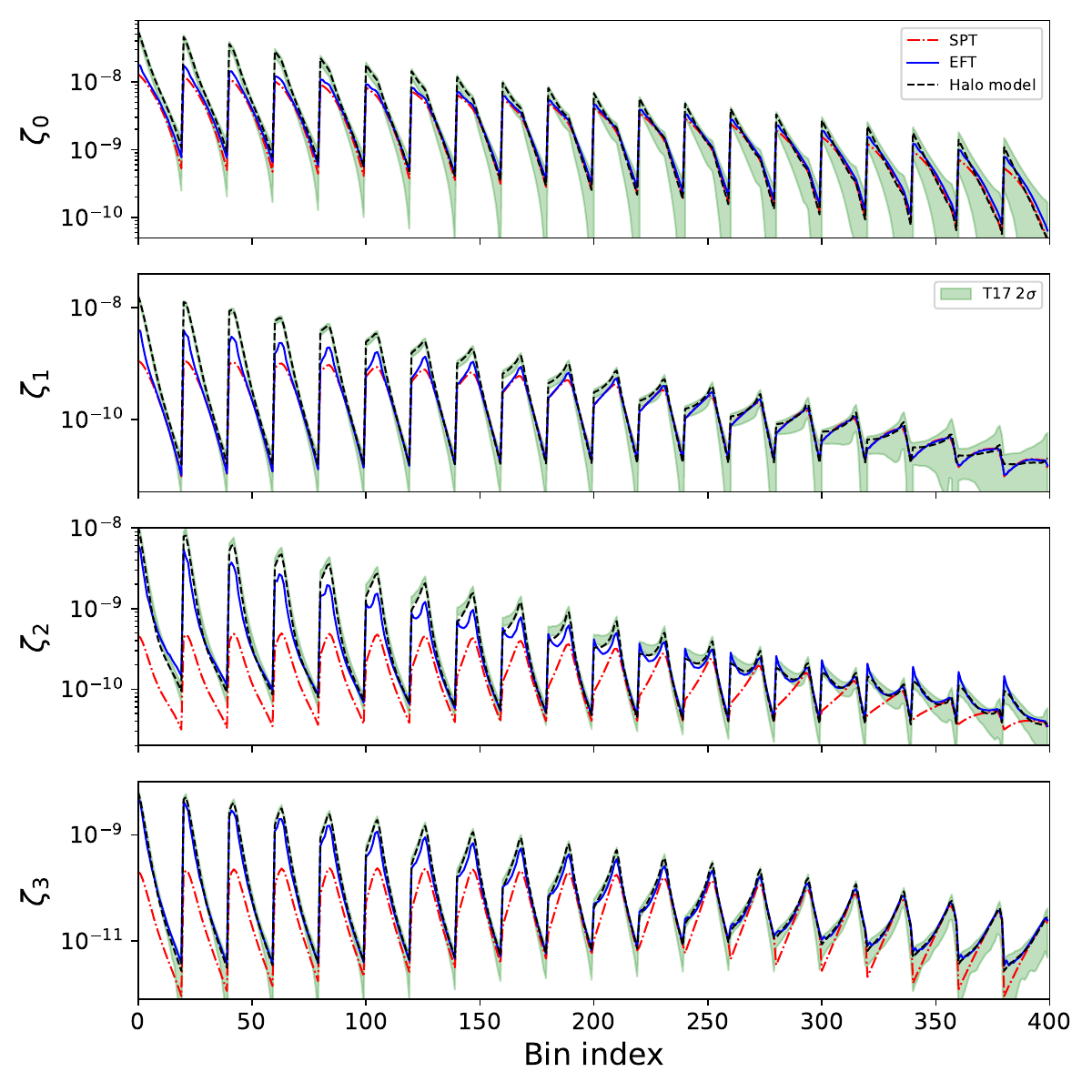}
	\caption{Bin index plots showing the performance of each model in single 1-dimensional graphics. We display the halo model (black dashed lines), EFT (blue lines), and SPT (red dot-dashed lines). The green dashed regions show the $2\sigma$ intervals around the mean of the 108 T17 realizations for the redshift bin centred at $z=0.5$.}  \label{fig:flatten_models}
	\end{center}
\end{figure*}

In this section we validate the performance of the models we have developed in the previous sections. First, in figure \ref{fig:squareszs9} we show 2-dimensional density plots of the correlation function coefficients $\zeta_m(\theta_1,\theta_2)$ for the first four multipoles $m=0$, 1, 2, 3. Here, we display the signal of the T17 simulations obtained with our code \texttt{cBalls}, as well as the phenomenological Halo model and theoretical EFT and SPT models. 

For the EFT model we have chosen 
\begin{equation} \label{eq:alphaD}
 \alpha^\text{EFT}(t) = \alpha_0 D_+^2(t).   
\end{equation}
The precise temporal evolution of this equation is intricate and analytically elusive. Nonetheless, considering that counterterms are derived from 2-point correlators of the small-scale fields dropped out from the theory (e.g. \cite{Carrasco:2012cv}), it can be expected that they evolve, at least, with the square of the growth function $D^2_+$. While the exact time dependence typically hinges on the scaling of the power spectrum \cite{Pajer:2013jj}, its accurate determination has only been approximately computed through simulations \cite{Foreman:2015uva}. Consequently, alternative, more generalized dependencies may be chosen instead. Our choice of scaling is motivated by the fact that this way, the entire correction to the linear power spectrum scales as $ \alpha^\text{EFT}(t) k^2 P_L(k,t)  \propto D_+^4(t)$, just as 1-loop contributions evolve.

Our modeling code takes about one minute to run a single cosmology, hence we are not yet in position to fit the simulations directly with sampling parameter algorithms. After empirically trying a few values, we choose the constant of proportionality in eq.~\eqref{eq:alphaD} to be $\alpha_0 = 3 \, h^{-2}\, \text{Mpc}^{2}$. 

\begin{figure*}
	\begin{center}
	\includegraphics[width=6 in]{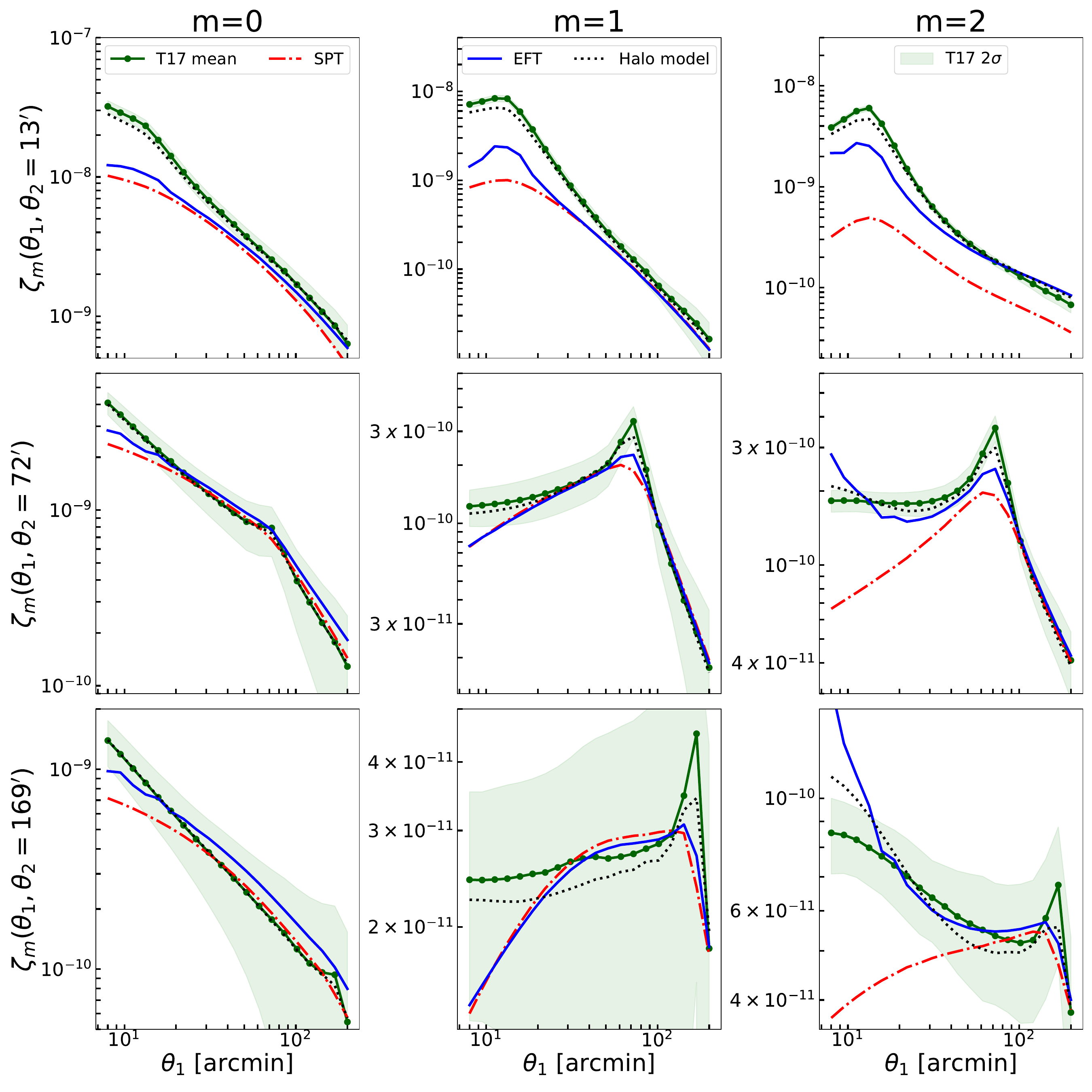}		
	\caption{1-dimensional plots for the 3PCF multipoles $\zeta_m(\theta_1,\theta_2)$ with $\theta_2$ fixed. We show the halo model, EFT and SPT. We use the T17 full-sky simulations for the redshift bin $z=0.5$. The means of T17 are depicted in red dots joined by straight lines and the pink shadows show the $2\sigma$ error regions. }  \label{fig:methods}
	\end{center}
\end{figure*}

In figure \ref{fig:flatten_models} we display a Bin index plot, in which the first index corresponds to the pair $(\theta_1^1,\theta_2^1)=(8',8')$ and the last one to the pair $(\theta_1^1,\theta_2^1)=(200',200')$, where the 2-dimensional space in, for example, the top panel of fig.~\ref{fig:squareszs9} is spanned from the bottom-left to the top-right bins. In these plots the light green shadings show the $2\sigma$ regions around the mean of the 108 T17 simulations for the bin centre at redshift $z=0.5$. We further display the models obtained with SPT in red dot-dashed lines, EFT in blue solid lines, and the \texttt{BiHalofit} model with black dashed lines. Panels from top to bottom show the multipoles $m=0,1,2,3$. Each 20 indices, the $\zeta_m$s returns to the smallest scale in one of its arguments. Hence, one expects higher accuracy in the models just before these jumps, particularly for higher indices, since in this case the two arguments of the 3PCF multipoles are the largest. We observe that the three models capture well the simulations patterns at large scales. However, SPT poorly fails at intermediate and small scales. The halo model is quite accurate at all scales, while adding the EFT correction show notable improvements over SPT. 

Complementing the graphics in figure \ref{fig:flatten_models}, we show figure \ref{fig:methods}, where we fix one of the arguments of the $\zeta_m$ multipoles ($\theta_2$) to different values and plot the whole range of $\theta_1: \,(8,200)\,\text{arcmin}$. In these plots the patterns are more cleanly depicted. Particularly the peak structure when the lines hit the diagonal $\theta_1=\theta_2$ is very well followed for all the modelings. This diagonal corresponds to all the isosceles triangles where the third size ($\theta_3$) can be arbitrarily small, hence one expects non-linearities to be important at these points and hence a more clear failure of the models. As expected, this is particularly true when the arguments are small. Another aspect that is worthy to note, is that the richness of the patterns of the 3PCF multipoles is followed very well for all the modelings, particularly, the SPT improvement accounted with the introduction of the $\alpha$ parameter in the EFT model is remarkable. However, the halo model stands out as the best-performing model, closely matching all the data extracted from the T17 simulations across all scales.

\begin{figure*}
	\begin{center}
	\includegraphics[width=5.4 in]{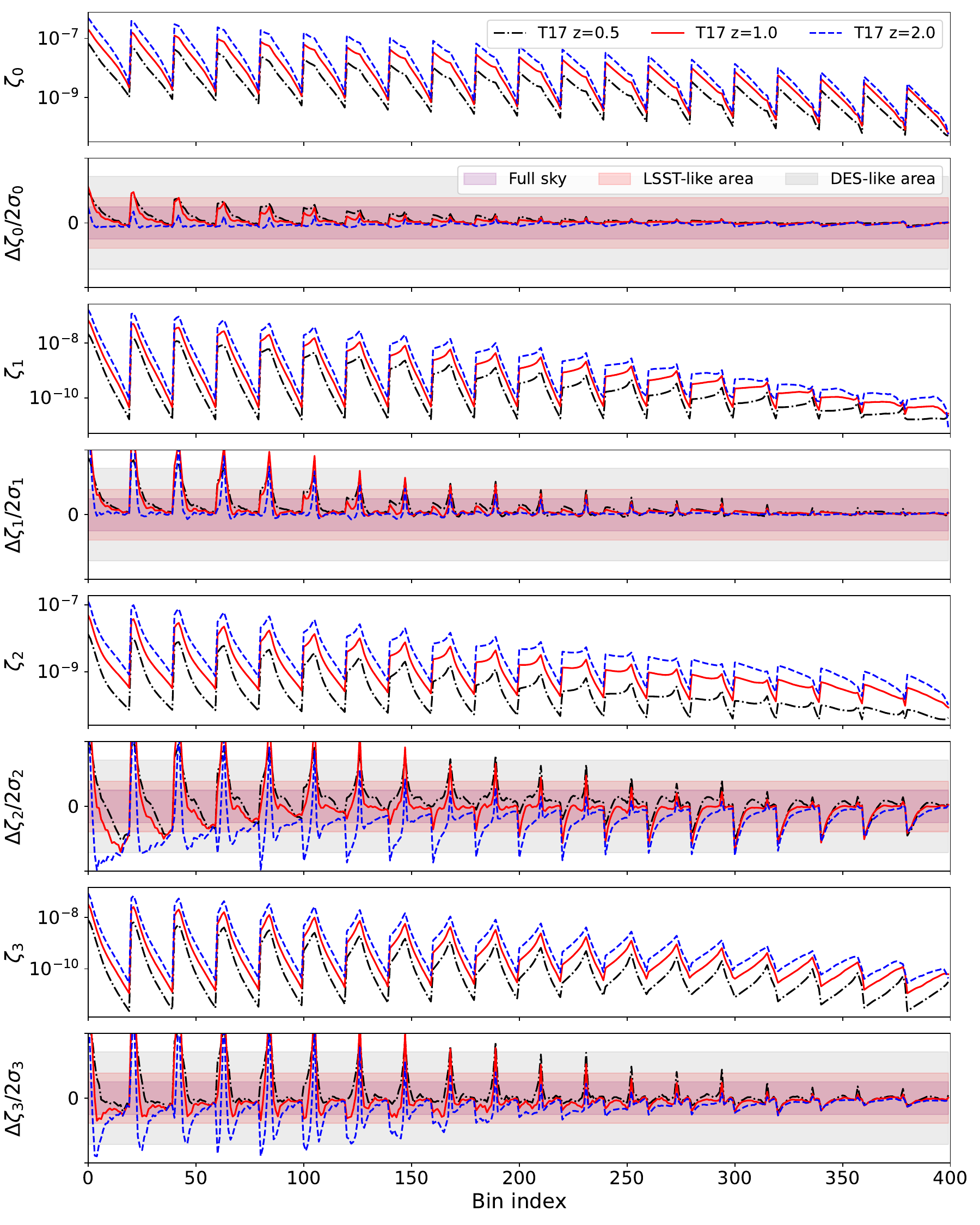}
	\caption{Bin index plots for redshift bins $z=0.5, 1.0$ and $2.0$. From top to bottom we show the signals given by the halo model followed by the plots showing if this falls inside the error $2\sigma$ intervals for the full-sky simulations (inner bands). We also depict the errors for the expected areas covered by LSST (middle bands) and DES (outer bands), under the simplification that the covariance matrix is Gaussian and by re-scaling the dispersion using the fractional volume of each experiment. Instead of re-scaling the signal for each sky coverage, we scale the y-axis for each sky fraction so that the corresponding band always stretches from $-2\sigma\leq\Delta\xi\leq +2\sigma$.}\label{fig:redshifts}
	\end{center}
\end{figure*}

In addition to the redshift bin $z=0.5$, we also compute the correlations at the bins $z=1.0$ and $z=2.0$. 
This is presented in figure \ref{fig:redshifts}, which show the different redshifts within the bin index structure described above. The lower panels show if the modeled signal falls within a 2$\sigma$ interval around the mean of T17. That is, 
\begin{equation}
    \Delta \zeta_m =  \zeta_m^\text{T17} -\zeta_m^\text{Halo model},
\end{equation}
with $\sigma$ the standard deviation of the data at that bin. In the same figure, we show not only the Halo model but also for the SPT and EFT models, taking into the account the error of the signal. The reference error band is defined by $-1<\Delta \zeta_m / (2 \sigma)<1$, which corresponds to the 2$\sigma$ limits of the full-sky simulations. \revised{To give us a very rough estimation of real surveys, we also show the equivalent bands for areas covered by DES and LSST (DES-like and LSST-like areas, respectively)}, constructed from the fractional sky covered with respect the full sky. These are computed by assuming a Gaussian covariance matrix approximation. That is, we rescale the associated errors by the factor $(A_\text{Full-Sky}/A_\text{experiment})^{1/2}$, where the area of the sphere (full-sky) is $A_\text{full-sky}=41,253\, \text{deg}^2$. Using $A_\text{DES}=5,000\, \text{deg}^2$  we obtain the middle band corresponding to the DES experiment, while using $A_\text{LSST}=18,000\, \text{deg}^2$ we obtain the outer band corresponding to LSST. Most of the modelled bins lie within the error bands, except for those corresponding to the peaks in the signal. The discrepancy between the model and the signal arise at the smallest scales, which correspond to the periodic jumps and also a leftward shift in the horizontal scale of the bin index plots of figure \ref{fig:redshifts}). However, for the quadrupole and higher multiples, there is an additional departure between the modeling and signal when $\theta_1=\theta_2$. 
A part of this departure comes from the previously discussed over counting of our estimator, but most of it is due to the non-linear nature of isosceles triangles when one scales tends to zero in a given triangle configuration. A similar conclusion has been reached for galaxy counting \cite{Slepian:2016weg}.

\end{section}

\begin{section}{Discussion and prospects}

Higher order statistics are compelling tools to extract non-Gaussian information of the matter field in the large scale structure of the Universe. Moreover, using these statistical tools in combination with the 2PCF have a number of benefits, including stronger constraints of the cosmological parameters, breaking degeneracies of observational and instrumental systematics, probing new physics in the gravitational or the particle sectors, reducing the number of feasible inflationary models through primordial non-Gaussianities, among others. However, the use of these higher order correlations should be accompanied by an accurate modeling of their signal, which is the main purpose of this work for the particular case of the 3PCF over projected scalar fields. Although, cosmological PT naturally resides in Fourier space, the are a few arguments that suggest working in real space may be more beneficial. For example, it is easier to deal with the observational window function. A recent discussion \cite{Baumann:2021ykm}, also suggest that it may be easier to decouple primordial non-Gaussianities from the non-linear evolution due to their local behaviour in position space. Therefore, we focus on the modeling of correlation functions and not their Fourier space counterparts.

In this work, we present a theory for the 3PCF decomposed on a plane wave expansion for scalar fields defined on the sphere, employing the tangent plane approximation. Building upon previous work by several authors \cite{Szapudi:2004gg,Zheng:2004eh,Pan:2005ym,Slepian:2015qza,Philcox:2021eeh}, we develop an estimator for measuring the multipoles, $\zeta_m$, of the 3PCF in this framework. The obtained computational complexity scales as the 2PCF with the data volume, substantially reducing the time required to compute the 3PCF for large datasets, such as those that will be obtained the by Vera Rubin Observatory over the 10 year operation period.

We use a C-code implementation of the multipole based estimator, named \texttt{cBalls}, which extracts the 3PCF coefficient signals in a reasonable amount of time for over 200 million \texttt{HEALPix} pixels in the weak lensing convergence maps of the full-sky T17 simulations \cite{Takahashi:2017hjr}. The analysis of such maps reveals a rapid decay of the multipoles $\zeta_m$ amplitude as $m$ increases, which could potentially enable one to reconstruct the complete 3PCF signal if desired. However, we argue that this is an indication that most of the information is likely concentrated in the first few multipoles, hence pursuing large $m$ coefficients may be unnecessary. Actually, we argue in favour of using the multipoles as summary statistics by themselves, without the need to compute the whole 3PCF. Considering the $\zeta_m$ as projections of the entire $\zeta$ correlation function brings out deep questions regarding the cosmological significance of high multipoles. These queries, including their degree of independence and the utility in cosmology, could be addressed through Fisher forecasts, which we leave for future work.

We develop analytical models to characterize the signal of the 3PCF multipoles, which then we contrast with the T17 simulated data. On a first approach, we formulate an SPT-based model, which shows good agreement with the data at large scales only (beyond roughly 60-100 arcminutes in both triangle sides and depending on the multipole $m$). Adding short-range of non-locality corrections \cite{McDonald:2009dh}, or EFT terms as we name them in this work, notably enhances the agreement with the simulations across a broader range of scales. A third and final approach to the modeling is to use the halo model which, as expected, shows the best performance when comparing to data, exhibiting remarkable agreement across nearly all scales tested, ranging from 8 to 200 arcmin.

Even though the halo model describes better the signal, the SPT approach offers insights into the role of the non-linearities. In the perturbative analytic description, we observe that the kernels to calculate the non-linear density fluctuations can be pulled out of the line-of-sight integral which describes the weak lensing convergence field from the matter field. This feature holds to all orders in PT, and it is the consequence of not introducing new scales into the theory. Actually, this fact remains true not only for extensions to general relativity which also have kernels that depend on the wave-number quotients only, but also in situations where the EdS kernels remain a good approximation, as it happens with massive neutrinos which indeed introduce a new scale. As long as $\Omega_m^{0.5} = f$ is a good approximation, the EdS kernels are reliable. For example, the errors of using EdS kernels in $\Lambda$CDM is below the percent in loop corrections. We expect similar results for dark energy models close to $\Lambda$CDM. Moreover, the presence of EFT terms do not change this fact. In summary, one may calculate the line-of-sight integrals over the linear power spectra of a given theory, and then applied the loop kernels to construct {\it N}-point correlations beyond the tree-level.

Although the method in this work is applicable to any projected scalar field over the sphere, we focus on the weak lensing convergence signal. For this particular field, coefficients of the 3PCF expansion, $\zeta_m(\theta_1,\theta_2)$, exhibit a pattern wherein two distinct power laws intersect at the points where $\theta_1$ equals $\theta_2$. This peculiar structure becomes apparent when examining 1-dimensional log-log plots with one of the variables $\theta$ maintained constant. Extending this work to model non-scalar fields, such as shear\footnote{An initial estimator for shear was developed by \cite{Porth:2023dzx} without a modeling of the signal. However, during the internal reviewing process of this publication by the DESC collaboration, a model for the shear multipoles appeared in \cite{Sugiyama:2024uqo}.}, should be easily achieved with our methods, but we leave those studies to future publications. Additionally, the modelling presented here might be useful to study systematics in the traditional 3$\times$2pt analysis. For example, one may wonder if adding 3pt statistic to the analysis of \cite{Leonard:2024nnw} can break some of the degeneracies found between photo-z errors and IA models. As such, the combination of 2pt with 3pt statistics may also prove useful to understanding other systematics in weak lensing, such as recovering redshift information (e.g.\cite{Guandalin:2021sxw}), blending or the dependence on a fiducial cosmology.

One may also wonder how fast and accurate the modeling of the convergence field is to perform a cosmological analysis for the upcoming LSST weak lensing observations. In terms of precision, the modeling described here matches the signal for most of the triangular configurations in the three redshift bins (0.5, 1 and 2.0) within the expected uncertainty of the LSST, with better accuracy towards smaller redshifts. Moreover, a further benefit of pulling the kernels out of the line-of-sight integrals is that the 2-dimensional Hankel transforms reduce to two 1-dimensional integrals, which then can be speedily estimated using the standard FFTLog methods. As a result, the modeling is efficient enough to be used for cosmological parameter inference, although one may also think of using emulators to reduce the computational time. Finally, this is just a first step towards using the 3PCF multpoles in the weak lensing of LSST, as well as other survey data, and further analyses should be carried out to understand the gain and challenges that it posses.

\end{section}

\acknowledgments

The authors would like to thank Oliver Philcox for discussions and advice in the early stage of this work, as well as Joachim Harnois-Deraps and Mike Jarvis for reviewing the manuscript in the DESC-LSST internal publication process. We also acknowledge Cora Uhlemann, Boryana Hadzhiyska, Jiamin Hou and Felipe Andrade-Oliveira for fruitful discussions. 

AA, AA, GN, MARM, JCH, SS and EM acknowledge support by CONAHCyT CBF2023-2024-162. AA (Abraham), GN, EM also acknowledge the support of DAIP-UG, and are grateful for the computational resources of the DCI-UG DataLab. AA (Alejandro), JCH and SS also  acknowledge support by PAPIIT IG102123. The authors acknowledge the LSST-MX Consortium for the managerial to facilitate their participation in the Rubin Observatory (fisica.ugto.mx/~lsstmx/). AA and MARM also acknowledge support by CONAHCyT CBF2023-2024-589. 

MARM~acknowledges that the analysis of T17 simulations in this work used the DiRAC@Durham facility managed by the Institute for Computational Cosmology on behalf of the STFC DiRAC HPC Facility (www.dirac.ac.uk). The equipment was funded by BEIS capital funding via STFC capital grants ST/K00042X/1, ST/P002293/1, ST/R002371/1 and ST/S002502/1, Durham University and STFC operations grant ST/R000832/1. DiRAC is part of the National e-Infrastructure in the UK. 

The DESC acknowledges ongoing support from the Institut National de Physique Nucl\'eaire et de Physique des Particules in France; the Science \& Technology Facilities Council in the United Kingdom; and the Department of Energy, the National Science Foundation, and the LSST Corporation in the United States.  DESC uses resources of the IN2P3 Computing Center (CC-IN2P3--Lyon/Villeurbanne - France) funded by the Centre National de la Recherche Scientifique; the National Energy Research Scientific Computing Center, a DOE Office of Science User Facility supported by the Office of Science of the U.S.\ Department of Energy under Contract No.\ DE-AC02-05CH11231; STFC DiRAC HPC Facilities, funded by UK BEIS National E-infrastructure capital grants; and the UK particle physics grid, supported by the GridPP Collaboration.  This work was performed in part under DOE Contract DE-AC02-76SF00515.

The contributions from the primary authors are as follows. AA (Abraham): tested the accuracy of the CBalls code, obtained its benchmarks and did comparisons with treecorr. EM and SS: debugged the CBalls code, tested its accuracy and ran most of the science comparisons between the theory and simulations. MARM: main developer of the CBalls code, and supervisor of junior members AA and EM  to test the performance and accuracy of the code. JCH: supervised the junior member SS in her work, reviewed the draft, and laboriously advised on the results and the general scope of the project. AA (Alejandro): project co-leader, main developer of the theory based on previous work, writer of the draft, supervisor of junior members SS and AA in their contributions to this project. GN: project co-leader, writer of the draft, supervisor of junior members EM and AA in this work.

\noindent We acknowledge the use of the following libraries and software: gsl, FFTW, numpy, scipy, matplotlib, Wolfram-Mathematica.

\appendix

\begin{section}{FFTLog for the line-of-sight integral} \label{sect:FFTLOG3pts}

We expand the linear power spectrum as a sum of scale invariant spectra with complex powers as \cite{TALMAN197835,Hamilton:1999uv}
\begin{align} 
\bar{P}_L(k) &= \sum_{m=-N/2}^{N/2} c_m k^{\nu + i \eta_m}, \label{pklfft} 
\end{align}
with phases
\begin{align}
\eta_m &= \frac{N-1}{N}\frac{2 \pi m}{\ln(k_\text{max}/k_\text{min})},    \label{pklfft_eta}
\end{align}
where we have split an interval $[k_\text{min},k_\text{max}]$ in $N$ logarithmic spaced  wave-numbers. The coefficients $c_m$ comes from the  discrete log-Fourier transform 
\begin{align} \label{cm}
c_m =  W_m k_\text{min}^{-\nu - i \eta_m} \frac{1}{N}\sum_{l=0}^{N-1} P_L(k_l)\left(\frac{k_l}{k_\text{min}}\right)^{-\nu} e^{- 2 \pi i m l/N},
\end{align}
with the weights $W_m =1$, except for the end points, for which $W_{-N/2}=W_{N/2} =1/2$.
The so-called bias $\nu$ is in principle an arbitrary real number, but its value is chosen to have a better convergence of the integrals \cite{Hamilton:1999uv}.

We need to solve integrals as
\begin{align} 
  I^{(n_1,m_1)}_{(n_2,m_2)}(\theta_1,\theta_2) 
  &=   \int d\chi \frac{q_A(\chi)q_B(\chi)q_C(\chi)}{\chi^4}  \left[\int \frac{\ell_{1} d\ell_{1}}{2 \pi}\ell^{m_{1}} P_\delta(\ell_{1}/\chi) J_{n_{1}}(\theta_{1} \ell_{1}) \right] \nonumber\\
 &\quad \times  \left[\int \frac{\ell_{2} d\ell_{2}}{2 \pi}\ell^{m_{2}} P_\delta(\ell_{2}/\chi) J_{n_{2}}(\theta_{2} \ell_{2}) \right],
\end{align}
where we allow for potential different photo-z distributions $A$, $B$ and $C$. 
We have to compute $A_0=I^{(0,0)}_{(0,0)}$, $A_1=I^{(1,1)}_{(1,-1)}+I^{(1,-1)}_{(1,1)}$ and $A_2=I^{(2,0)}_{(2,0)}$.\footnote{Note $A_1(\theta_1,\theta_2)=I^{(1,1)}_{(1,-1)}(\theta_1,\theta_2) + I^{(1,1)}_{(1,-1)}(\theta_2,\theta_1)$, since $I^{(1,-1)}_{(1,1)}(\theta_1,\theta_2)=I^{(1,1)}_{(1,-1)}(\theta_2,\theta_1)$}

Using $k_1=\ell_1/\chi$ and $k_2=\ell_2/\chi$, 
\begin{align} 
  I^{(n_1,m_1)}_{(n_2,m_2)} (\theta_1,\theta_2) 
  &=   \int d\chi \, q_A(\chi)q_B(\chi)q_C(\chi) \chi^{m_1+m_2} D_+^4(\chi) \left[\int \frac{dk_{1}}{2 \pi}k_{1}^{m_{1}+1} P_L(k_{1}) J_{n_{1}}(\theta_{1}\chi\, k_{1}) \right] \nonumber\\
 &\quad \times   \left[\int \frac{dk_{2}}{2 \pi}k_{2}^{m_{2}+1} P_L(k_{2}) J_{n_{2}}(\theta_{2}\chi\, k_{2}) \right].
\end{align}

We permit different biases $\nu_{1}$, $\nu_{2}$ when $m_1\neq m_2$, hence we write
\begin{align}
    \bar{P}_L(k_1) &= \sum_{s_1=-N/2}^{N/2} c_{s_1} k_1^{\nu_{b_{m_1}} + i \eta_{s_1}},\\ \bar{P}_L(k_2) &= \sum_{s_2=-N/2}^{N/2} c_{s_2} k_2^{\nu_{b_{m_2}} + i \eta_{s_2}}
\end{align}
then
\begin{align} 
  I^{(n_1,m_1)}_{(n_2,m_2)}(\theta_1,\theta_2) 
  & = \sum_{s_1=-N/2}^{N/2} \sum_{s_2=-N/2}^{N/2} c_{s_1} c_{s_2}  \int d\chi \, q_A(\chi)q_B(\chi)q_C(\chi) \chi^{m_1+m_2} D_+^4(\chi) 
  \nonumber\\ &\quad \qquad \times   
  \left[\int \frac{dk_{1}}{2 \pi}k_{1}^{a_{s_1}}  J_{n_{1}}(\theta_{1}\chi\, k_{1}) \right] \left[\int \frac{dk_{2}}{2 \pi}k_{2}^{a_{s_2}}  J_{n_{2}}(\theta_{2}\chi\, k_{2}) \right]
\end{align}
with $a_{s_1}= 1+b_{m_1}+i \eta_{s_1}$ and  $a_{s_2}= 1+b_{m_2}+i \eta_{s_2}$.

Now, we use the integral
\begin{align}
    \int \frac{dk }{2 \pi}\; &k^{m+a_s}  J_{n}(x k)  =\frac{2^{-1+m+a_s} x^{-m-a_s-1} \Gamma \left(\frac{1}{2} (m+a_s+n+1)\right)}{\pi \Gamma \left(\frac{1}{2} (-a_s-m + n + 1)\right)},
\end{align}
and obtain

\begin{align} 
  I^{(n_1,m_1)}_{(n_2,m_2)}(\theta_1,\theta_2) 
  & = \sum_{s_1=-N/2}^{N/2} \sum_{s_2=-N/2}^{N/2} c_{s_1} c_{s_2}  A_{s_1 s_2} \theta_1^{-(1+a_{s_1}+m_1)}  \theta_2^{-(1+a_{s_2}+m_2)} \nonumber\\
  &\quad \times \frac{2^{-1+m_1+a_{s_1}} \Gamma \left(\frac{1}{2} (1+m_1+a_{s_1}+n_1)\right)}{\pi \Gamma \left(\frac{1}{2} (1-a_{s_1}-m_1+n_1)\right)}  \nonumber\\
  &\quad \times
  \frac{2^{-1+m_2+a_{s_2}} \Gamma \left(\frac{1}{2} (1+m_2+a_{s_2}+n_2)\right)}{\pi \Gamma \left(\frac{1}{2} (1-a_{s_2}-m_2+n_2)\right)},
\end{align}
with

\begin{equation}
    A_{s_1 s_2} = \int d\chi \, q_A(\chi)q_B(\chi)q_C(\chi)  \chi^{ -a_{s_1} -a_{s_2}-2} D_+^4(\chi).
\end{equation}

The matrix elements $A_{s_1 s_2}$ can be stored, while the matrix multiplications can be performed very fast using linear algebra computational packages.


\end{section}

 \bibliographystyle{JHEP}  
 \bibliography{refs.bib}

\end{document}